\documentclass[prr,aps,twocolumn,superscriptaddress]{revtex4-2}
\usepackage{amsmath, amsthm, amssymb,float}
\usepackage{amsfonts}
\usepackage{graphicx}
\usepackage{dcolumn}
\usepackage{bm}
\usepackage{textcomp}
\usepackage[normalem]{ulem}

\usepackage[titletoc,title]{appendix}

\usepackage{color}
\begin{document}

\title{On the capillary discharge in the high repetition rate regime}

\author{P.~Sasorov} \email{pavel.sasorov@eli-beams.eu}
\affiliation{Extreme Light infrastructure ERIC, 252 41 Dolni Brezany,  Czech Republic}
\author{G.~Bagdasarov} 
\affiliation{Extreme Light infrastructure ERIC, 252 41 Dolni Brezany,  Czech Republic}
\author{N.~Bobrova}
\affiliation{Faculty of Nuclear Sciences and Physical Engineering, Czech Technical University in Prague, Brehova 7,
115 19 Prague, Czech Republic}
\author{G.~Grittani}
\affiliation{Extreme Light infrastructure ERIC, 252 41 Dolni Brezany,  Czech Republic}
\author{A.~Molodozhentsev}
\affiliation{Extreme Light infrastructure ERIC, 252 41 Dolni Brezany,  Czech Republic}
\author{S.V.~Bulanov} 
\affiliation{Extreme Light infrastructure ERIC, 252 41 Dolni Brezany,  Czech Republic}
\affiliation{National Institutes for Quantum and Radiological Science and Technology (QST),
Kansai Photon Science Institute, Kizugawa, Kyoto 619-0215, Japan}

\begin{abstract}
We investigate the main physical processes that limit the repetition rate of capillary discharges used in laser accelerators of electrons theoretically and with computer simulations. We consider processes in the capillary. We assume that a cooling system independently maintains  temperature balance of the capillary, as well as a gas supply system and a vacuum system maintain conditions outside the capillary.
 The most important factor, determining the highest repetition rates in this case, is the capillary length, which governs a refilling time of the capillary by  the gas. For a short capillary, used for acceleration of sub-GeV electron beams, the repetition rate approximately equal to 10 kHz, which is inversely proportional to the square of the capillary length. The effects of the capillary diameter, gas type and the gas density are weaker. 
 
\end{abstract}

\maketitle

\nopagebreak

\section{Introduction}

Compact laser-based accelerators of relativistic electrons conceive a broad
variety of applications, including compact free electron lasers (FELs)~\cite{Gr07,Fu09,Hu12,Co14,Ma12,Mo18,As20}, Compton sources~\cite{Ph12,Ge15,Kh15}, and electron-positron colliders~\cite{Le09,Sch10}. Refs.~\cite{Fa04,Ma04,Os08,Li11,Le14,Gon19} demonstrate  electron beam generation with laser wake-field acceleration (LWFA) mechanism~\cite{Ta79,Es09}.

Capillary discharges provide electron density distribution, which is suitable for guiding laser pulse driver, used in the LWFA scheme. The capillary discharges were discussed in details in Refs.~\cite{Ho00,Bu02,Sp03,Ka07,Kam09,Go11,Le14,Gon19,Pa50,Ti15,Po19,Eh96,Sp01,Bob02,Pie20,Bob13,Bag21}.

For many applications high  repetition rate ($\gtrsim 1$~kHz~) of the electron accelerator operation is required~\cite{Ge15}. A hydrogen-filled capillary discharge waveguide for LWFA, operating at kHZ repetition rates  is considered in Ref.~\cite{Gon16}. Other plasma waveguides for LWFA  operating at the high repetition rates are described in Refs.~\cite{Al22,Arc22}. Such plasma channels can be used to obtaini multi-GeV electron beams~\cite{Le14,Gon19,MS22}.

In this paper we present the results of simulations and theoretical investigations of the high repetition rate operation and its dependence on the plasma waveguide parameters. We consider the following stages of a cycle: filling of the capillary by a cold gas, dynamics of the discharge from its initiation till almost complete recombination of the plasma, and expansion of the gas-plasma into the vacuum region and the gas supplying slots. We also take into account driver laser pulse energy deposition into capillary plasma. We neglect the effects of the processes outside the capillary discharge, leading to potential limitation to the repetition rate of capillary discharge, considered in Ref.~\cite{Gon16}.

We use the code MARPLE~\cite{Ga12} for simulations.

\section{Simulation setup}
\label{setup}

\begin{figure*}[ht] 
\includegraphics[width=0.8\textwidth,clip=]{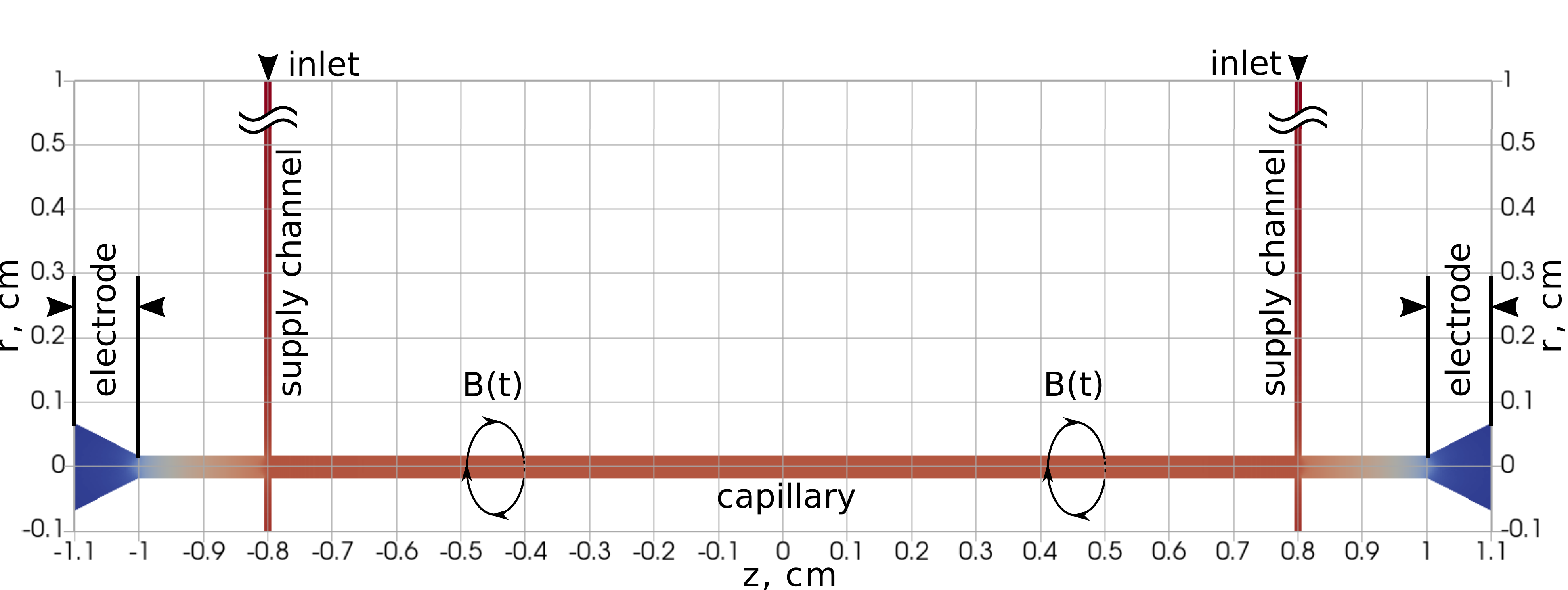}
\caption{Overall geometry of the simulation domain. It is azimuthally symmetric. It consists of: i) the capillary of  $D=0.34$~mm diameter and $L=2$~cm length within the region, $|z|\leq 1$~cm, with ideal rigid dielectric wall, ii) conical orifices in ideal metal electrodes  at $1.1~\mbox{cm}\leq|z|<1$~cm, and iii) two disk-like supply channels in ideal rigid dielectric with inlets  at $r=1$~cm .
}
\label{fig1}
\end{figure*}

Simulations of the capillary discharge, operating in a repetitive regime, are performed by use of the three-dimensional magnetohydrodynamic (MHD) code MARPLE. It implements two-temperature  (ion and electron) MHD equations. It takes also into account the electron and ion thermal conductions, viscosity and finite electric resistance. The physical model is described in Refs.~\cite{Ga12,Ol20,Bag22,Bag21}. This code has been used in simulations connected with various experiments~\cite{Gon19,Bag21,Bag17,Gr08,Ale19}.

Full cycle of operation of the capillary discharge in repetitive regime consists of two phases. One of them corresponds to the phase of the electric discharge, whereas the second one corresponds to dynamics of neutral hydrogen H$_2$ gas between the discharges. The cyclic operation of the capillary discharge starts from a process of filling the capillary with this gas. The both phases are simulated with the same code. However, simulation of the electromagnetic field as well simulation of electron component of the plasma are switched off during the phases of neutral hydrogen dynamics simulations. The phase of the gas discharge is simulated with the full MHD code described briefly above.

We assume a circular cross-section shape capillary of the length $L=2$~cm and diameter $D=2r_0=340\,\mu$m. 
The computational domain is shown in Fig.~\ref{fig1}. It includes two supply channels of the width $D/3$ at $|z|=0.8$~cm as well as the electrode regions ($1\,\mbox{cm}<|z|<1.1\,$cm). They play significant role in the gas and plasma dynamics inside the capillary~\cite{Bag17,Bag22}.

The considered configuration is azimuthaly symmetric. We assume that capillary is made of ideal absolutely rigid dielectric, whereas the electrodes are made of ideal absolutely rigid metals. These assumptions allow us  to set the boundary conditions on the discharge stage. Boundary conditions at $|z|=1.1$~cm provide free outflow of the  gas or plasma. We set the temperature of the capillary and the temperature of electrodes are equal to a room temperature. Simulating the discharge dynamics, we assume capillary wall and electrodes temperature is of
0.5~eV that is much lower than typical temperature of the discharge plasma. 

\begin{figure}[ht] 
\includegraphics[width=0.45\textwidth,clip=]{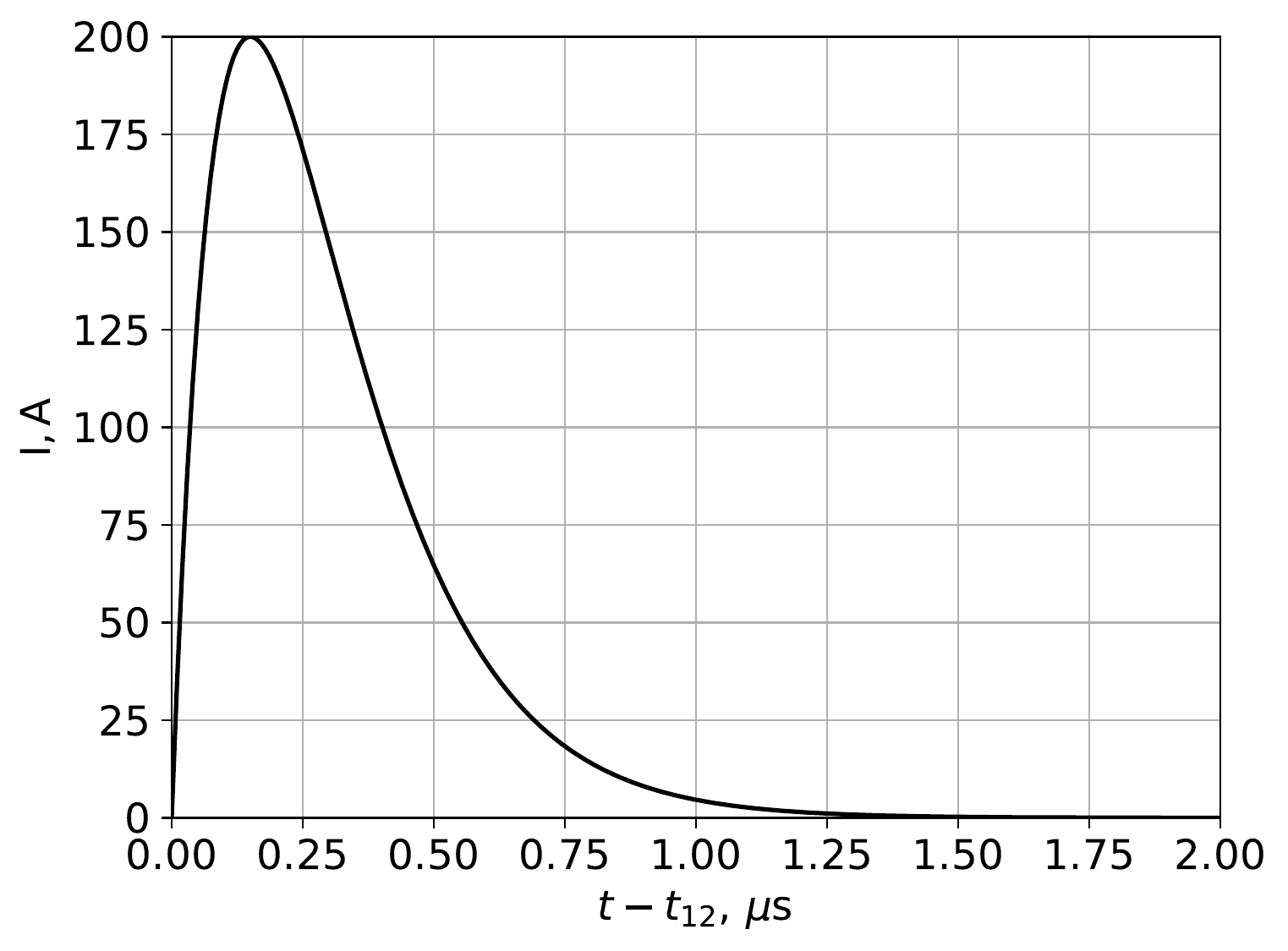}
\caption{Total electric current through the capillary discharge vs time during stage~2. 
}
\label{fig1a}
\end{figure}

The simulation is split into the three  stages, stages 1, 2 and 3. Stage 1 models the process of the capillary filling with a molecular hydrogen before the discharge is ignited. The hydrogen gas at the pressure $p=$75~mbar and at room temperature fills capillary through the supply channels (inlets marked in Fig.~\ref{fig1}). The gas pressure is chosen to achieve  electron density required for LWFA. The final almost steady state flow obtained at stage 1 (at $t=t_{12}=200\,\mu$s ) is used as an initial condition for the capillary discharge MHD simulations (stage~2). 
The discharge current has the following form:
\begin{equation}\label{curr}
I(t>t_{12})=I_0\frac{t-t_{12}}{t_c}\exp\left(1-\frac{t-t_{12}}{t_c}\right),
\end{equation}
where $I_0=200$~A and $t_c=150\,$ns. Its profile is shown in Fig.~\ref{fig1a}. The azimuthal component of magnetic field at the insulator boundary is set to be $B_\varphi=2I(t)/(cR)$, where $I(t)$ is given by Eq.~(\ref{curr}), and  $R$ is the current radius at the boundary of the computational domain. This determines the boundary condition for magnetic field. At $t=t_{23}=t_{12}+2\,\mu\mbox{s}=202\,\mu$s, when discharge ended and plasma  cooled down, stage 2 is finished. Stage 3 models the process of recovering the neutral hydrogen distribution in the capillary, though some capillary refilling  begins  by the end of stage 2. Final state of stage 2 (at $t=t_{23}$) is used as the initial condition for stage 3. When steady-state density distribution is reached, stage 3 ends  at $t- t_{23}\sim 100$-150$\,\mu$s. Thus, the transition between stages 1 an 2  takes place at $t=t_{12}=200\,\mu$s, and the transition between stages 2 and 3 occurs at $t=t_{23}=202\,\mu$s.

The physical model implemented in the code MARPLE, presented above, describes evolution of gas-plasma density, $\rho$, electron and ion temperatures, $T_e$ and $T_i$, and radial and axial velocities, $v_r$ and $v_z$ inside the computational domain. The model describes also evolution of azimuthal component of magnetic field, $B_\varphi$. 
For stages 1 and 3 we exclude the magnetic field, $B_\varphi$, and electron temperature, $T_e$, from the system equations used in the physical model and set $T=T_i$.  All mentioned above values depend (in the azimuthally symmetric case) on $r$, $z$ and $t$, where $r$ is radial coordinate of the cylindrical coordinate system, $(r,\varphi,z)$, associated with the capillary axis. 

All main parameters of the three stage simulation, $a$, $L$, axial electron density $2\times 10^{18}\,\mbox{cm}^{-3}$, energy of the driver femtosecond laser pulse, 3~J, correspond to the project~\cite{Mo18}. We consider below the set of these parameters as nominal.

The same discretization of the computational domain is used in all simulation stages. Spatial step of $h_r=5\,\mu$m is used  along the capillary radius and  in the most interesting regions of the computational domain in z-direction, e.g. junction of supply channel and capillary.  The total amount of computational cells of 474K is partitioned into 112 pieces and processed in parallel.

\section{Results of  the simulation}

\subsection{Stage 1: Simulation of capillary filling}
\label{st1}

\begin{figure*}[ht] 
\includegraphics[width=0.78\textwidth,clip=, bb = 0 0 1500 440
]{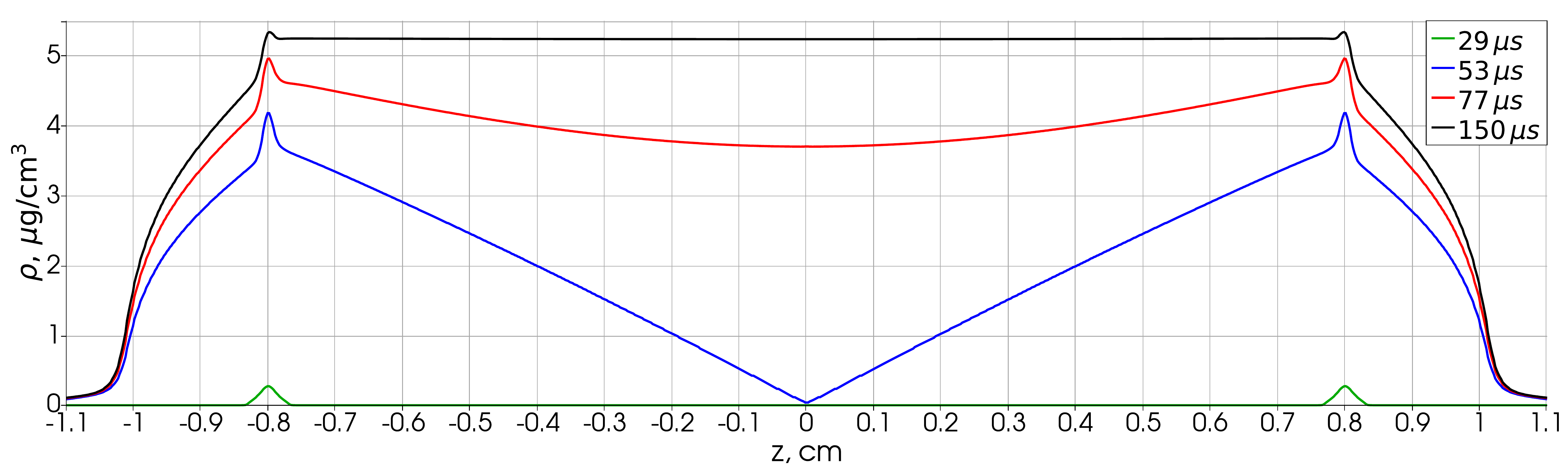}
\caption{Gas density distribution during the stage 1.  Shown is 1d density distributions at the axis of the capillary for different time moments. The curves are labeled by the time $t$. 
}
\label{fig2}
\end{figure*}

\begin{figure}[ht] 
\includegraphics[width=0.45\textwidth,clip=
]{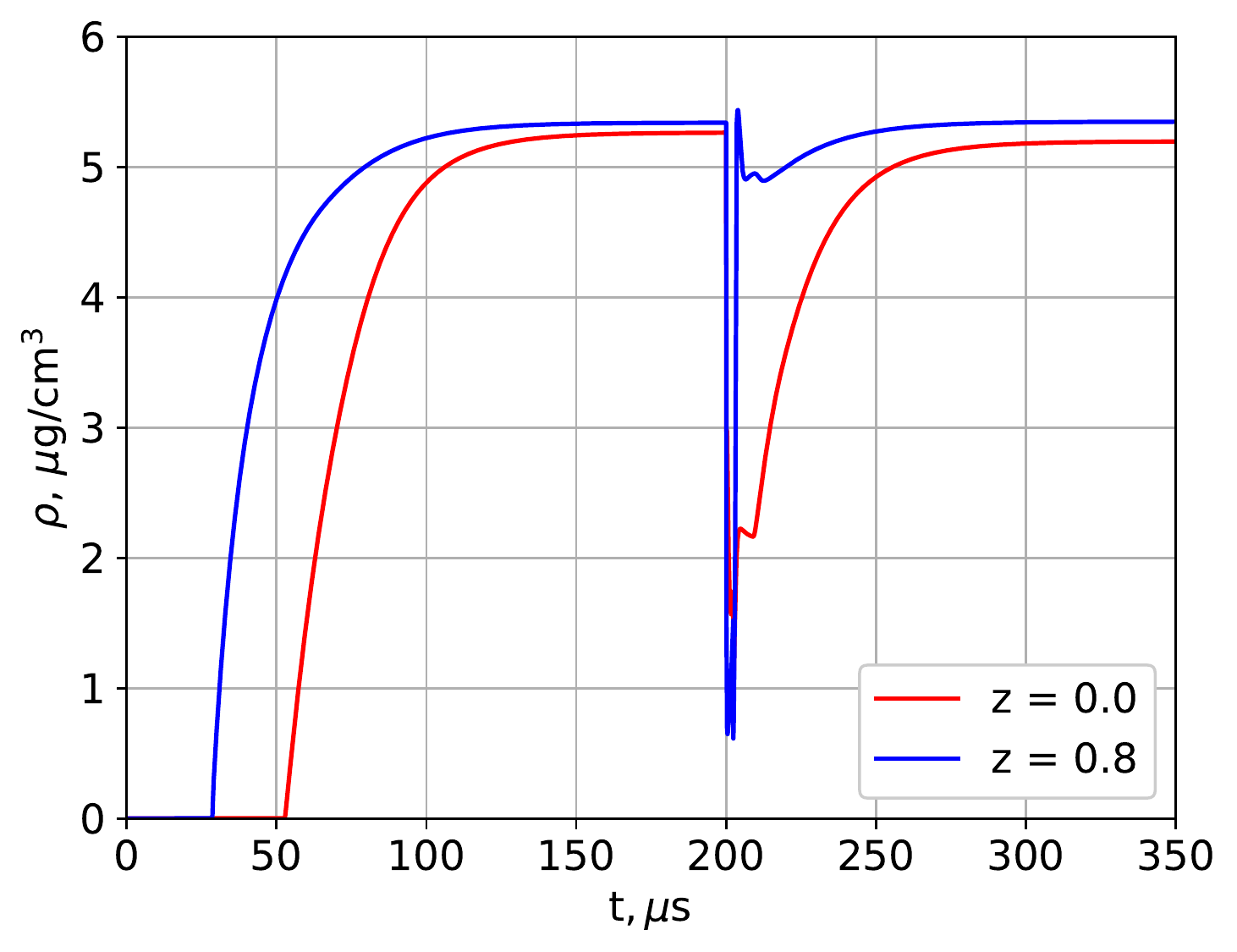}
\caption{The red line presents hydrogen density evolution at the center of the capillary ($r=0$, $z=0$) during the whole simulation: stages 1+2+3. The blue line shows the same but for $r=0$, $|z|=0.8$~cm, that are positions of the supply channels.  
}
\label{fig3}
\end{figure}

\begin{figure}[ht] 
\includegraphics[width=0.45\textwidth,clip=
]{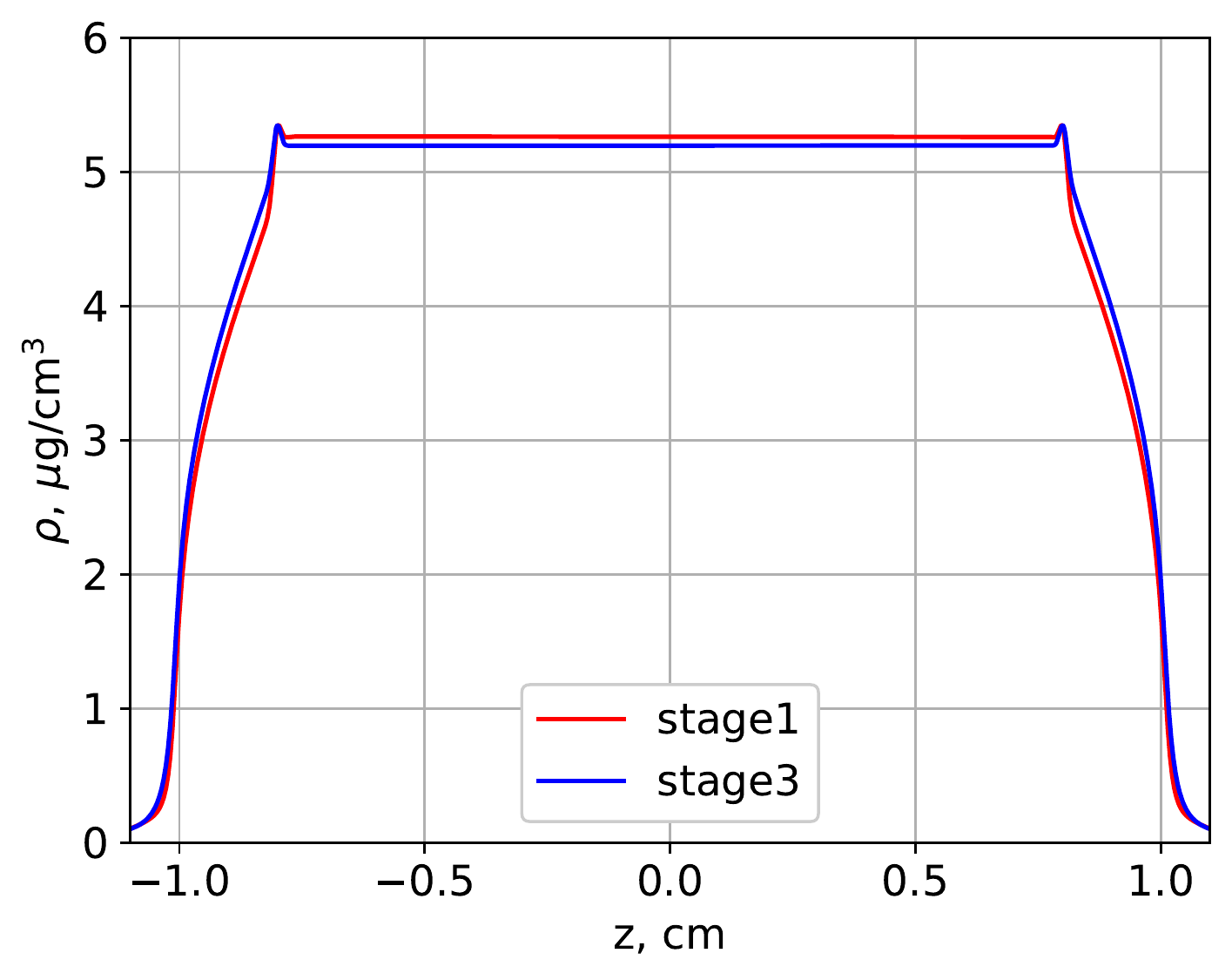}
\caption{Steady state distributions of axial gas densities at asymptotically long times at the stages 1 and 3. 
}
\label{fig9}
\end{figure}

\begin{figure}[ht] 
\includegraphics[width=0.45\textwidth,clip=]{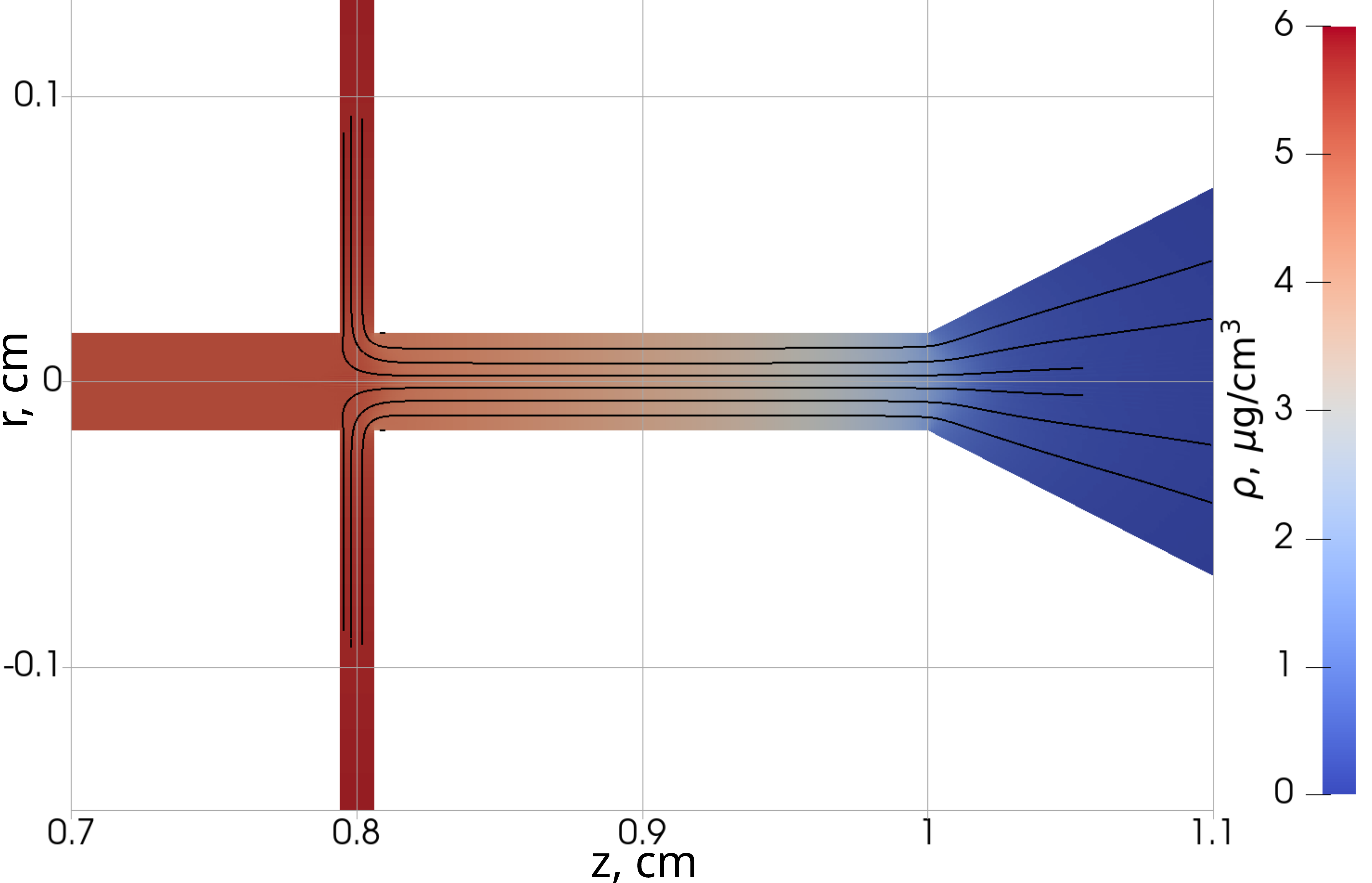}
\caption{2D hydrogen density distribution and lines of its velocity at the steady state of stage 1. A spatial domain that is not so far from the capillary end is shown only. In the region that is significantly to the left of the supply channel, the density is constant (see Fig.~\ref{fig9}) and the gas is almost motionless (see Fig.~\ref{fig71}).
}
\label{fig70}
\end{figure}

\begin{figure}[ht] 
\includegraphics[width=0.45\textwidth,clip=]{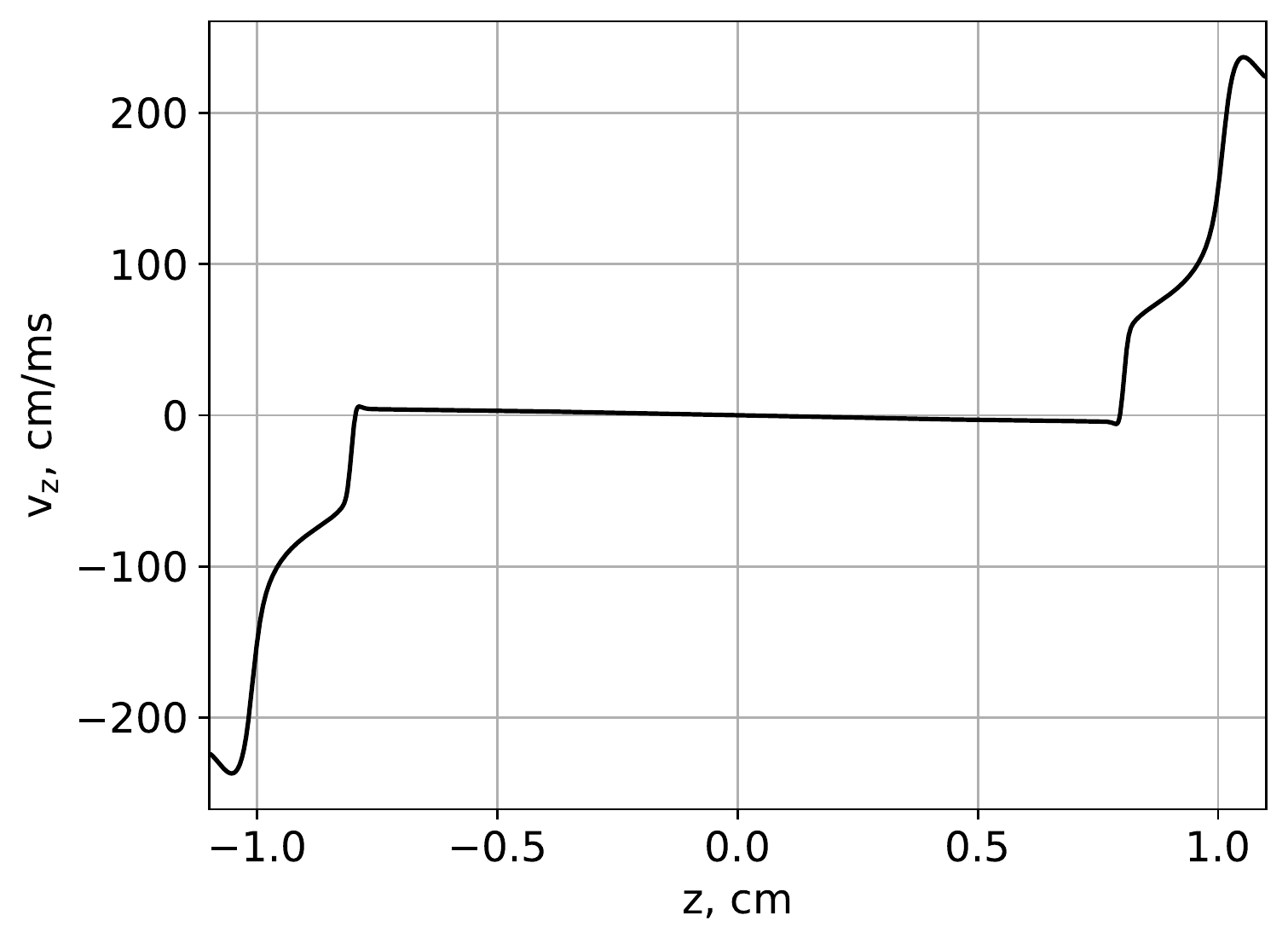}
\caption{Steady state axial velocity, $v_z(0,z,t_{1\infty})$, vs $z$ for stage1. }
\label{fig71}
\end{figure}

 The main goal of this stage is to obtain a spatial distribution of the gas at the end of the process of capillary filling, when a steady state gas flow of is established.
The evolution of the hydrogen density distribution is shown in Fig.~\ref{fig2}. Fig.~\ref{fig3} shows density time evolution  at the center of the capillary at $r=0$, $z=0$ for all three stages of the simulation. The time interval $0<t<t_{12}=200\,\mu$s  corresponds to stage 1.   Due to the thermal conductivity gas temperature  is almost equal to the capillary wall temperature, 300$^\circ$~K, during this stage. Figs.~\ref{fig2} and~\ref{fig3} show that the hydrogen flow approaches the steady state. The steady state, shown in Figs.~\ref{fig9}, \ref{fig70} and~\ref{fig71}, is established at the time $t_{1\infty}=200\,\mu$s. Fig~\ref{fig9} shows axial density versus $z$, Fig.~\ref{fig70} shows the two-dimensional gas density distribution in the vicinity of the supply channels and  the gas velocity field lines.  Fig.~\ref{fig71} shows axial velocity distribution along the capillary axis. Figs.~\ref{fig70} and~\ref{fig71} allow us to conclude, that the gas is almost motionless in the capillary between the supply channels. Existence of such steady state gas flow is very important for the problem of the repetition rate. Expense of the hydrogen in the steady state is equal to 0.14~mg/s per each channel. The process of the relaxation towards the steady state together with the analogous results for stage 3 determine a recovery time for the steady state density distribution.

When deviation of the central density, $\rho(0,0,t)$, from its steady state value, $\rho_\infty$ becomes less than 5\%,   temporal evolution of the central density is described by the expression:
\begin{equation}\label{Q010}
\rho(0,0,t)\simeq\rho_\infty \left(1-0.5 \exp\frac{t_{*}-t}{\tau_{\scriptsize{\mbox{sim}}}}\right)\,,
\end{equation}
where $\rho_\infty=\rho(0,0,t_{1\infty})=5.26\,\mu$g/cm$^{3}$, $\tau_{\scriptsize{\mbox{sim}}}=16.9\,\mu$s and $t_{*}=66.7\,\mu$s. This expression gives reasonable approximation for $\rho(0,0,t)$ at the center of the capillary for  time $t\gtrsim t_*$.

\subsection{Stage 2: Simulation of capillary  discharge}

We use gas parameters of the steady state gas flow, obtained in stage 1, as initial conditions for plasma dynamic simulation during the discharge stage. The considered  model cannot describe the initial electric breakdown of the cold hydrogen.  However, the
breakdown lasts a short time, only about 10~ns, and presumably does not affect the 
plasma discharge properties, considered in the present paper.  
 Simulation of the second stage  begins, when gas becomes weakly ionized. Below, we consider results of the discharge simulation.
 
The present capillary discharge simulation differs in several important aspects from the 1D-3D simulations, whose results are presented  in Refs.~\cite{Bob02,Bag17,Bag22}. Transverse dynamics of the capillary plasma, leading to formation of the plasma channel suitable for laser pulse guiding, is similar to considered  in Ref.~\cite{Bob02}, where 1D approximation has been used. However,  the 2D consideration shows that the lifetime of the short channel is determined  by  plasma outflow from the capillary ends, but not by   the electric pulse duration as in the one-dimensional or  the long capillary cases (Ref.~\cite{Bag17}).

\begin{figure*}[ht] 
\includegraphics[width=0.9\textwidth,clip=
]{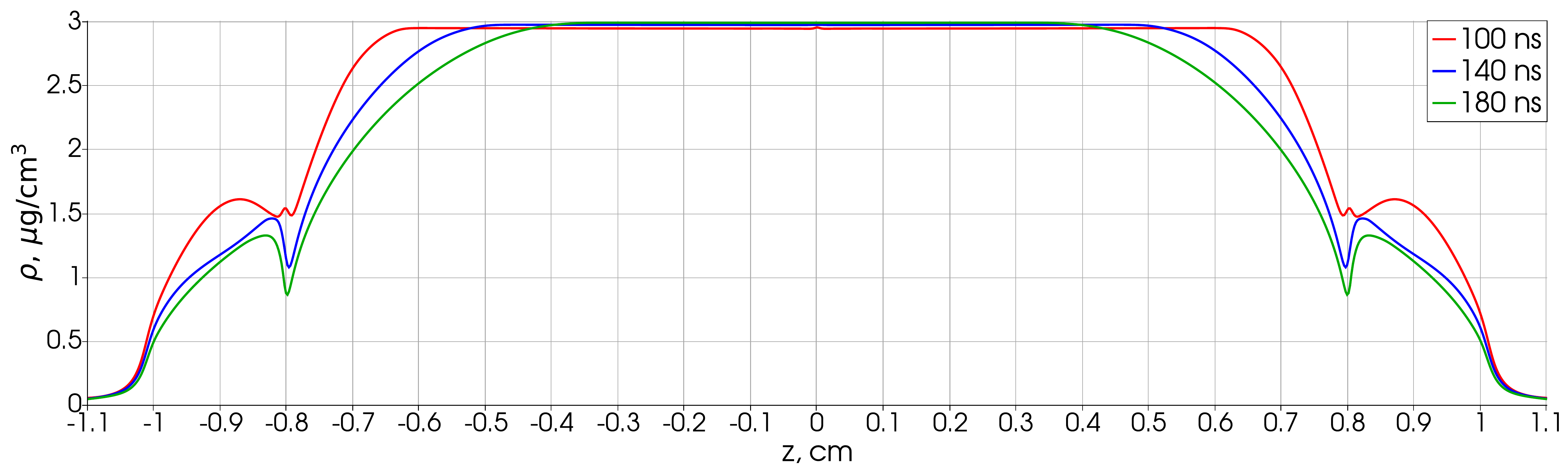}
\caption{Hydrogen density distribution during the stage 2. Only time interval of existing of well organized plasma channel of sufficient length  is presented here.
There are 1d density distributions at the axis of the capillary for different time moments. The curves are labeled by the time $t-t_{12}$. }
\label{fig4a}
\end{figure*}

\begin{figure}[ht] 
\includegraphics[width=0.45\textwidth,clip=]{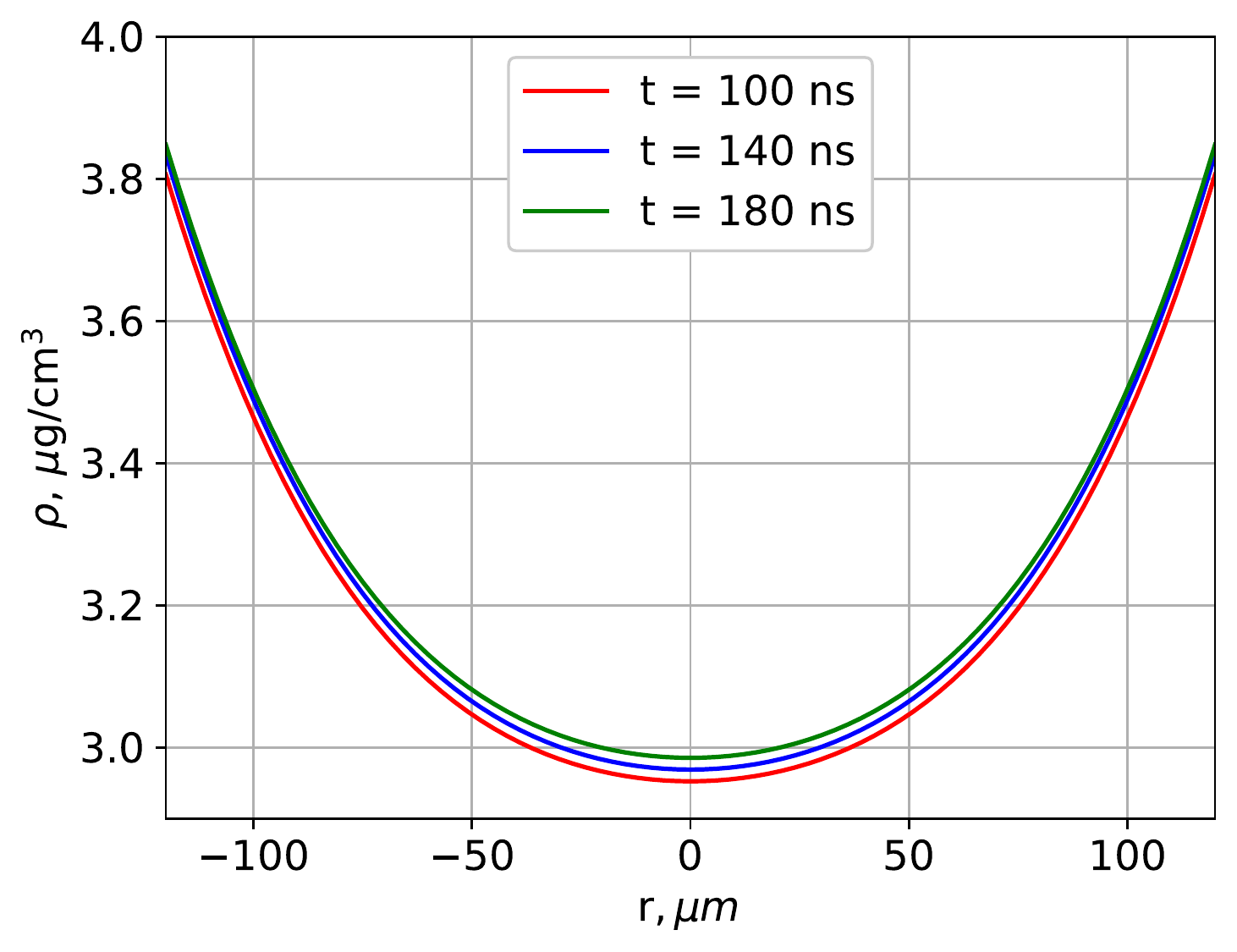}
\caption{Transverse plasma density distributions at $z=0$ at the same time moments as in Fig.~\ref{fig4a}. The curves are labeled by the time $t-t_{12}$. The plasma is almost completely ionized at these moments (see Fig.~\ref{fig4c}). Thus, these profiles represent also transverse electron density distributions. 
}
\label{fig4b}
\end{figure}

\begin{figure}[ht] 
\includegraphics[width=0.45\textwidth,clip=
]{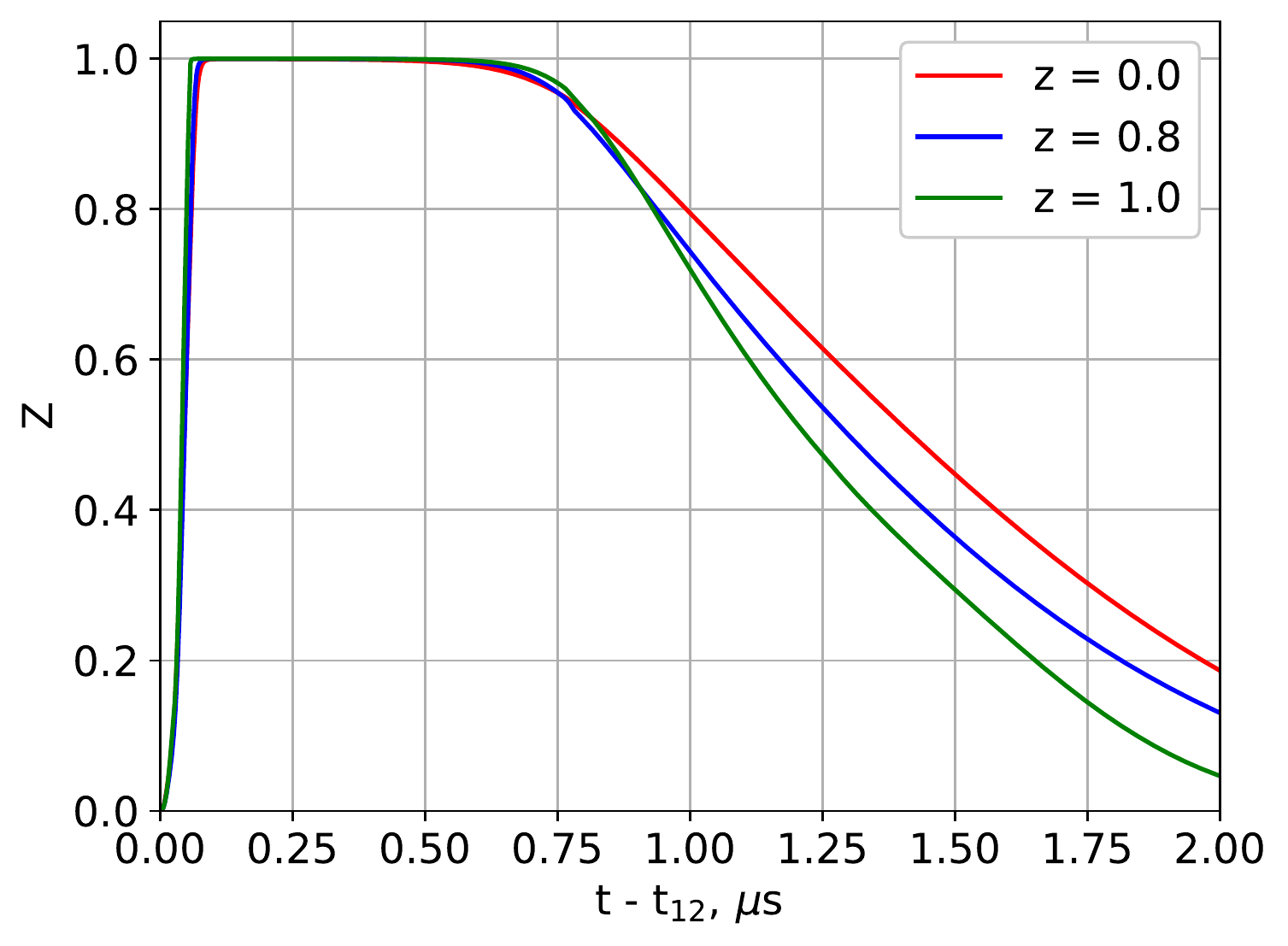}
\caption{Time dependencies of the axial mean ion charge $Z$ at several positions along the axis during stage 2, labeled by $z$. Well formed plasma channel for laser pulse guiding exists only while $Z$ is close to 1.}
\label{fig4c}
\end{figure}

Parameters of  the present simulation are similar to the case considered  in Ref.~\cite{Bag22}, where it was shown  that the capillary length (1 cm) is too short to provide sufficiently long axially homogeneous plasma channel. The plasma outflow into the supply channels shortened the effective length of axially uniform plasma channel. The latter effect was not  taken into account in the 2D $(r,z)$-simulation~\cite{Bag17}, because  total capillary length was much larger than  distance between  the supply channels and the capillary ends. The present capillary of 2 cm length is longer than the capillary considered in Ref.~\cite{Bag22}, other parameters are similar, hence during sufficiently long period, lasting about 80 ns, the length of axially homogeneous plasma channel is larger than 0.8~cm, as it is seen in Figs.~\ref{fig4a}, \ref{fig4b} and~\ref{fig4c}. Fig.~\ref{fig4a} shows distribution of plasma density at $r=0$ along the capillary axis for three time moments $t-t_{12}=100,\, 140$ and 180~ns. Fig.~\ref{fig4b} shows radial distribution of plasma density in the capillary center at $z=0$ for the same time moments. We see the plasma density minimum on the axis in the considered time interval. The plasma is completely ionized everywhere, excluding a thin layer of about $\sim 30\, \mu$m width near the capillary wall, during the time interval $\sim 80\mbox{~ns}<t-t_{12}<300$~ns.  Fig.~\ref{fig4c} shows time dependence of ionization degrees at different capillary cross sections. Electron density is proportional to the plasma density at least for the part of the capillary channel ($|z|\leq 0.4$~cm) during the time interval $t-t_{12}=100$-180~ns. Electron density axial value is almost constant and  approximately equal to $1.8\times 10^{18}$~cm$^{-3}$. The latter value fits the electron acceleration considered in Ref.~\cite{Mo18}. For  $0<t-t_{12}<180$~ns transverse plasma dynamics in this part of the capillary is described by the 1D model~\cite{Bob02}. Outside this region plasma density distribution is affected by the plasma outflow through the open ends of the capillary and  the gas supply channels.

Fig.~\ref{fig4a} shows  that  plasma channels longer than 1~cm exist for a shorter time, 
$100\mbox{~ns}<t-t_{12}<140$~ns.

\begin{figure}[ht] 
\includegraphics[width=0.45\textwidth,clip=
]{
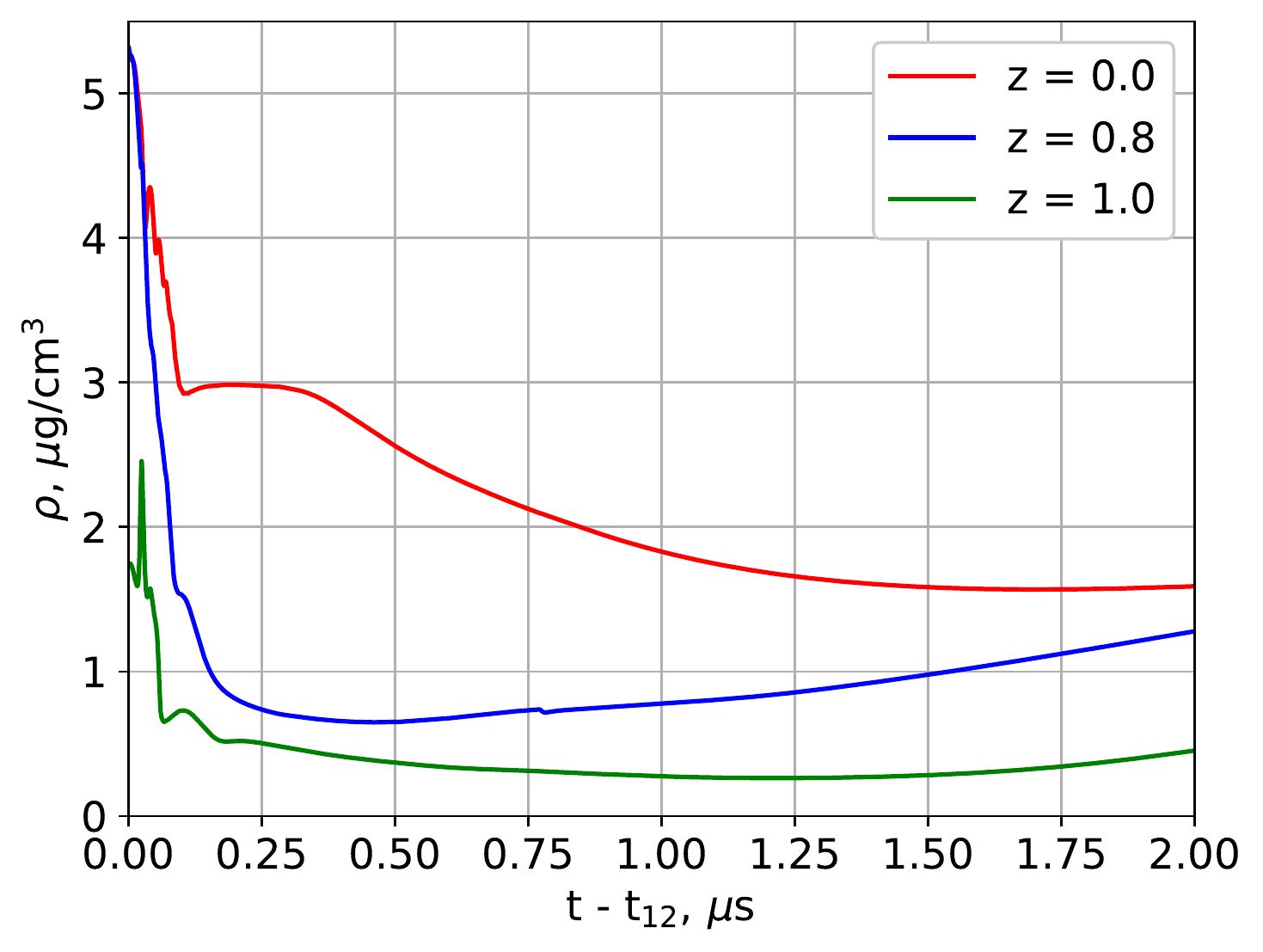}
\caption{Time dependencies of axial plasma density during stage 2 at  several  $z$.}
\label{fig5}
\end{figure}

\begin{figure}[ht] 
\includegraphics[width=0.45\textwidth,clip=
]{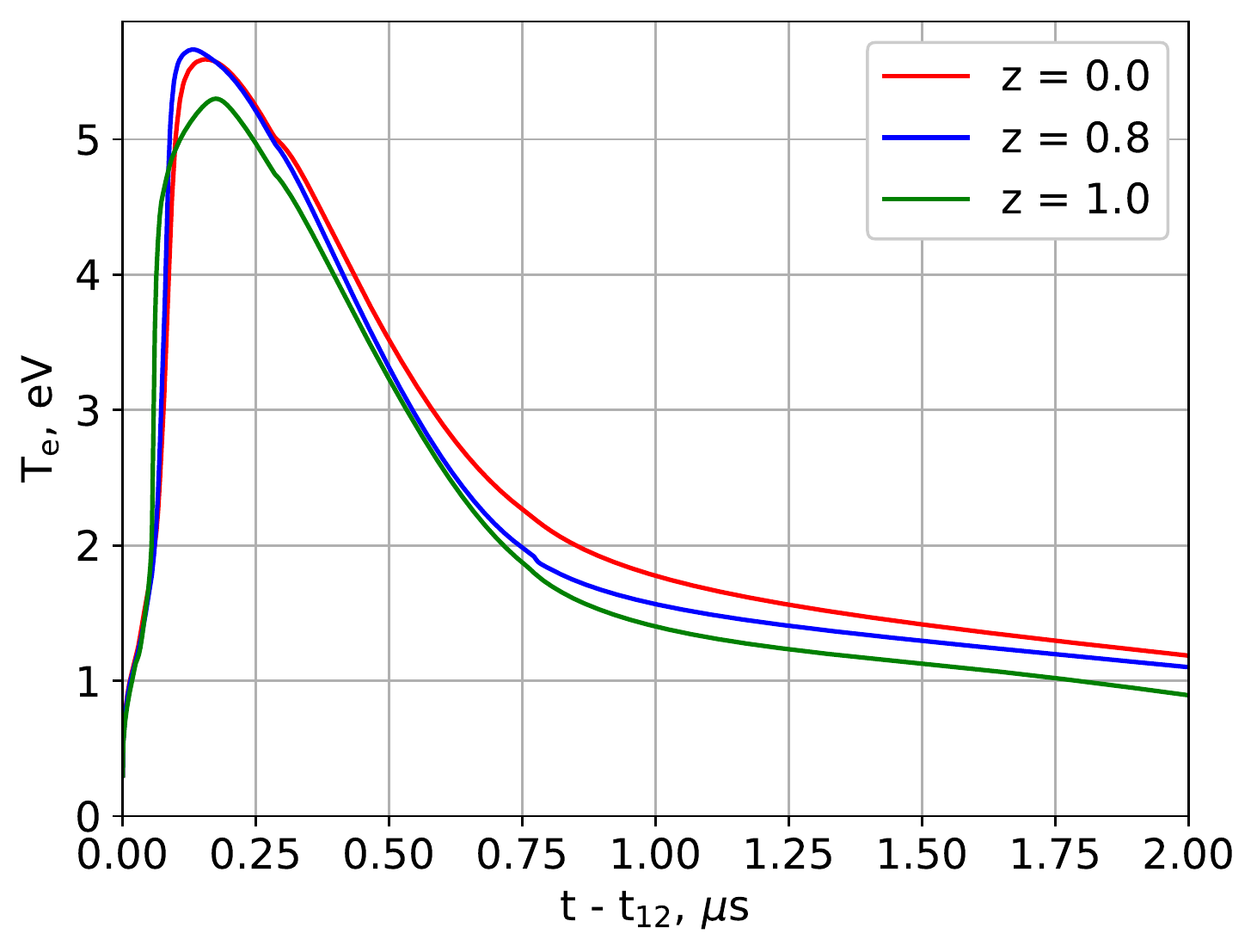}
\caption{Time dependencies of axial electron temperature during stage 2 at  several  values of $z$. 
}
\label{fig6}
\end{figure}

The present simulation of the capillary discharge is extended till the moment of almost complete recombination of hydrogen. The further simulation of neutral gas behavior in the capillary till the next pulse of electric current is considered below in Sec.~\ref{st3}. High plasma pressure during the discharge and right after it leads to an outflow of almost all hydrogen from the capillary. Fig.~\ref{fig5} shows plasma density on the capillary axis versus time at different capillary cross sections ($|z|=0$, 0.8 and 1~cm). Initial axial density decrease occurs during the beginning of the discharge ($t<80$~ns) and corresponds to radial redistribution of the plasma density~\cite{Bob02}. The redistribution leads to formation of  the electron density radial profile (see  Fig.~\ref{fig4b}), necessary for the laser guiding. The plasma temperature behavior during the discharge is shown in Fig.~\ref{fig6}. Further plasma dynamics  is determined mainly by plasma outflow from the capillary and by initial stage of its refilling. The gas-plasma pressure is diminishing due the gas-plasma outflow and plasma cooling  to the value, lower than the pressure inside the suppliers. The refilling starts approximately at $t-t_{12}\sim 500$-700~ns near the suppliers and at $\sim 1.25$-1.5~$\mu$s at the middle of the capillary. The gas-plasma outflow is caused by plasma heating during the discharge. This process takes place in time interval $t_{12}=200\,\mu\mbox{s}<t<t_{23}=202\,\mu$s in Fig.~\ref{fig3}.

The plasma outflow is accompanied by the plasma cooling  due to the  thermal conduction to the cold capillary walls  (see Fig.~\ref{fig6}) and, hence, by the recombination (see Fig.~\ref{fig4c}). After the time $t-t_{12}=2\,\mu$s the ionization degree becomes less than 0.2, and we end the  simulation of stage 2.

\subsection{Stage 3: Recovering of neutral hydrogen gas distribution after the plasma discharge}
\label{st3}

We start this  stage 3 of simulation  at the time $t=t_{23}=202\,\mu$s. The difference between stages 3 and 1  is as follows: i) initial gas temperature in the capillary is considerably higher than the wall temperature (see Fig.~\ref{fig6}); ii) initial gas density is not  zero, though  is significantly lower than the steady state density at the first stage; and iii) initially the supply channels are not empty. 

\begin{figure}[ht] 
\includegraphics[width=0.45\textwidth,clip=
]{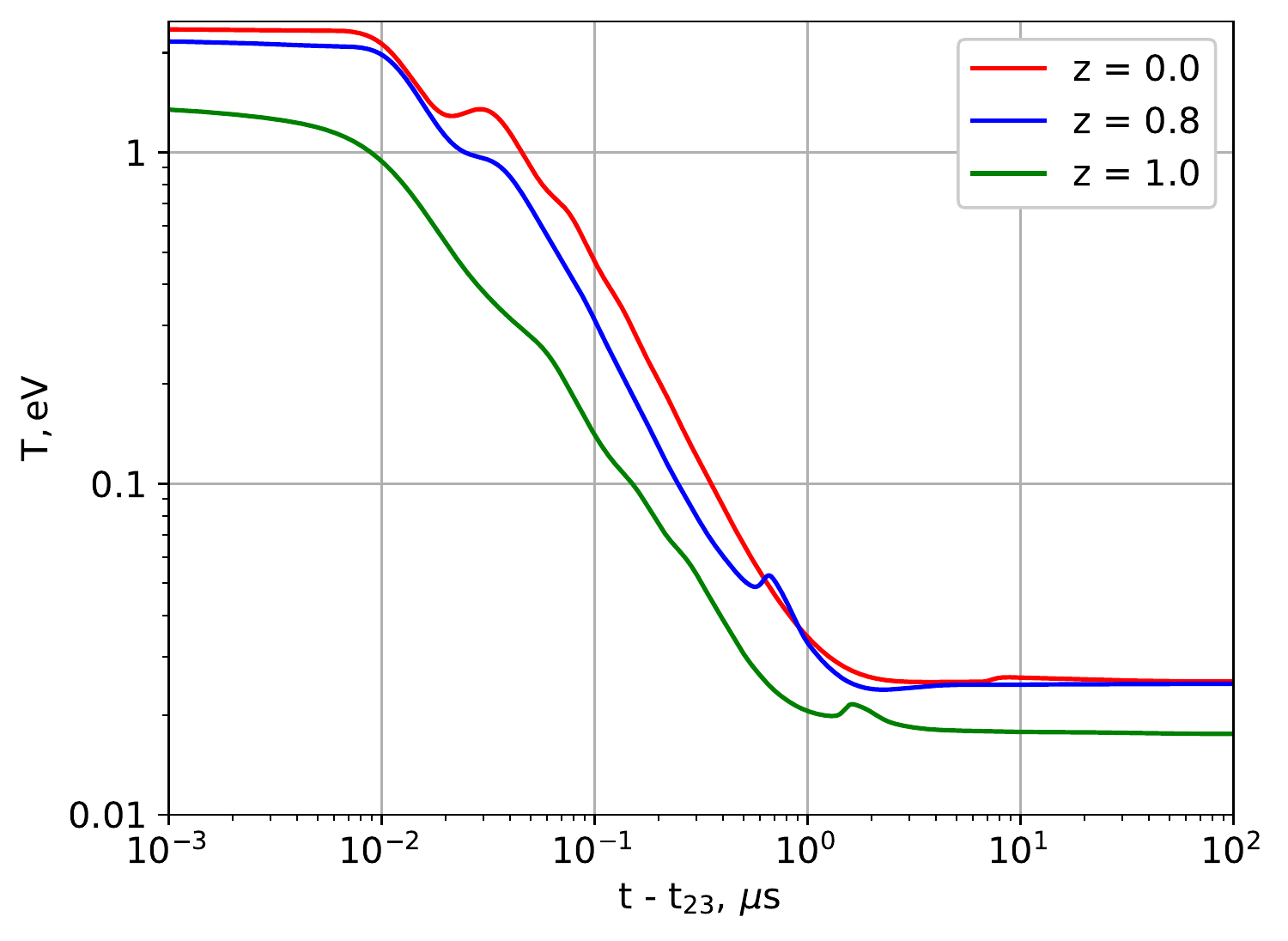}
\caption{Time dependencies of axial gas temperature during the beginning of stage 3 at different positions along the axis, labeled by values of $z$.}
\label{fig7}
\end{figure}

In the  simulation of gas-plasma dynamics during  stage 2 we use a plasma equation of state. This equation of state takes into account only atomic neutral particles. Simulating stage 3, we use equation of state, describing  neutral molecular hydrogen, H$_2$. These two different equations of states are not conjugated in the intermediate range of temperatures. So we cannot exclude discontinuities at the time of the transition from stage 2 to stage 3. We set the gas-plasma density and pressure to be continuous functions of time at every spatial point. This approach provides absence of any mechanical shocks at the transition. Nevertheless, the gas temperature has a discontinuity at the transition time, $t_{23}$. However, this leads to  negligibly weak change in the gas flow parameters during a few microseconds after the beginning of stage 3.  Fig.~\ref{fig7} shows that the temperature relaxes on the timescale of a few  microseconds to the capillary wall temperature. This relaxation time is considerably less than capillary refilling  characteristic time (100~$\mu$s). As a result, such temperature jump does not influence our main statements.
 
A similar problem  occurs  at the transition from stage 1 to stage 2. However, due to very high rate of energy deposition during the electric breakdown of hydrogen at the onset of the electric current pulse this period is very short and  introduces  much less inaccuracy into the simulation.

Stage 3 differs from stage 1  mainly due to the initial conditions. In both stages initial gas densities are much smaller than steady state ones. It is necessary to note that the initial densities in the supply channels are different. In stage 1 the initial gas density in the channels is equal to zero. In stage 3, it is of the order of the steady state density excepting the regions of the supply channels near the capillary. The geometry of the supply channels provides small deviation of the initial density distribution in the supply channels right after the discharge from a steady state distribution. Difference in the initial channel densities  leads to delay of the filling process by 60-70~$\mu$s in stage 1 compared to stage 3 (see Fig.~\ref{fig3}).

The initial gas temperatures are also different. It is approximately $25,000^\circ$~K in stage 3, and depends on $r$ and $z$. Then gas temperature quickly relaxes to the wall temperature. Fig.~\ref{fig7} shows that the relaxation process lasts about  few of $\mu$s. The capillary refilling time is about 100~$\mu$s. Comparing these times we conclude that the higher initial temperature in the stage 3 does not influence on the recovering process.

Steady state density distribution on the axis, obtained in the simulation at the end of stage 3, is shown in Fig.~\ref{fig9}. 
We consider the distribution at $t=t_{3\infty}=t_{23}+148\,\mu$s as  a steady state one. Relaxation of the  central  density, 
$\rho(0,0,t)$ towards  its steady state value, $\rho_\infty$, is shown in Fig.~\ref{fig3} (see also Fig.~\ref{fig8} below). When deviation of the central density, $\rho(0,0,t)$, from its steady state value, $\rho_\infty$ becomes less than 5\%, temporal evolution of the central density is described by the following expression:
\begin{equation}\label{Q020}
\rho(0,0,t)\simeq\rho_\infty \left(1-0.5 \exp\frac{t_{*}-t}{\tau_{\scriptsize{\mbox{sim}}}}\right)\,,
\end{equation}
where $\rho_\infty=\rho(0,0,t_{3\infty})=5.20\,\mu\mathrm{g/cm}^{3}$, $\tau_{\scriptsize{\mbox{sim}}}=17.5\,\mu$s and 
$t_{*}=t_{23}+8.0\,\mu$s.
We note that the parameters $\rho_\infty$ and  $\tau_{\scriptsize{\mbox{sim}}}$ are almost the same for the stages 1 and 3.
This expression describes the relaxation of $\rho(0,0,t)$  with a moderate accuracy up to the times $t=t_{*}$, when the  deviation from the steady state value becomes about $0.5\,\rho_\infty$.
At $t-t_{23}>100\,\mu$s, the relative difference of the density distribution inside the capillary from the steady distribution becomes well below 1\%.  So we estimate  the neutral hydrogen distribution recovery time after the discharge for considered capillary  parameters as 100~$\mu$s. It corresponds to the 10~kHz repetition rate. The problem of the repetition rate will be considered in more details below, in Sec.~\ref{rep}.

\subsection{Comparison of stages 1 and 3}
\label{comp}

\begin{figure}[ht] 
\includegraphics[width=0.45\textwidth,clip=
]{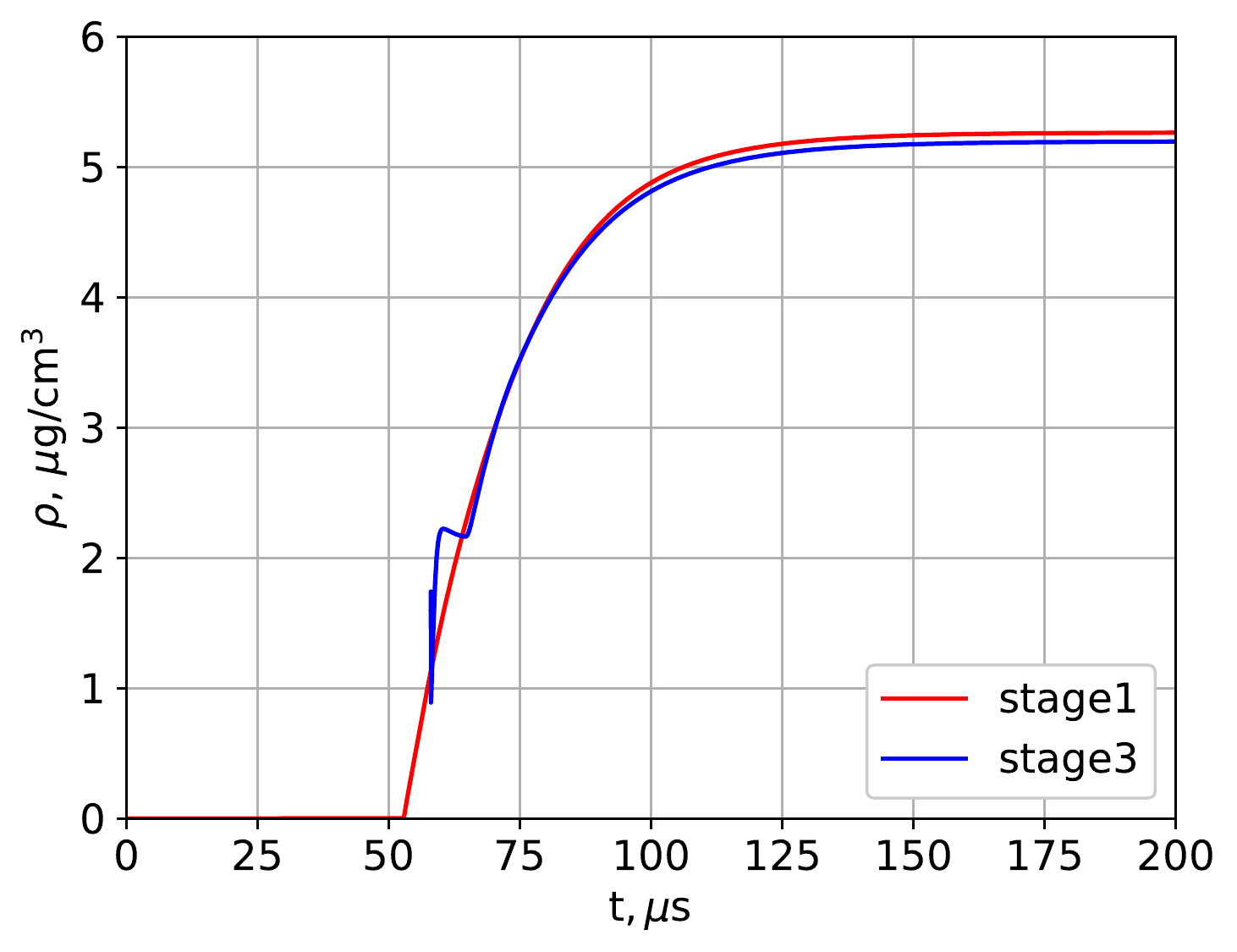}
\caption{Comparison of the gas density relaxation processes at the capillary center ($r=0$, $z=0$) for stages 1 and 3. The plot for stage 3 is shifted to the left by 144~$\mu$s for more convenient comparing.
}
\label{fig8}
\end{figure}

\begin{figure}[ht] 
\includegraphics[width=0.45\textwidth,clip=
]{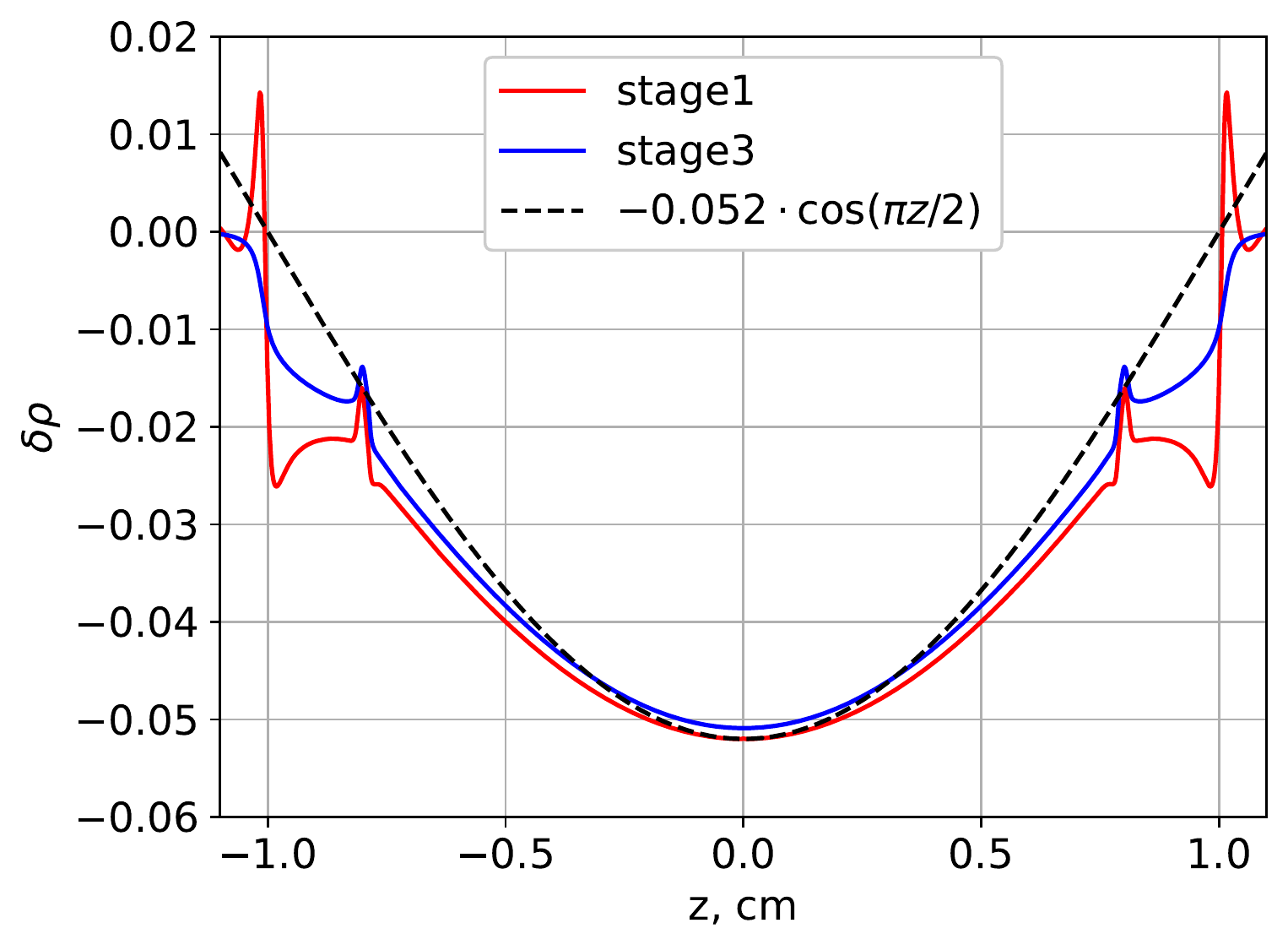} 
\includegraphics[width=0.45\textwidth,clip=
]{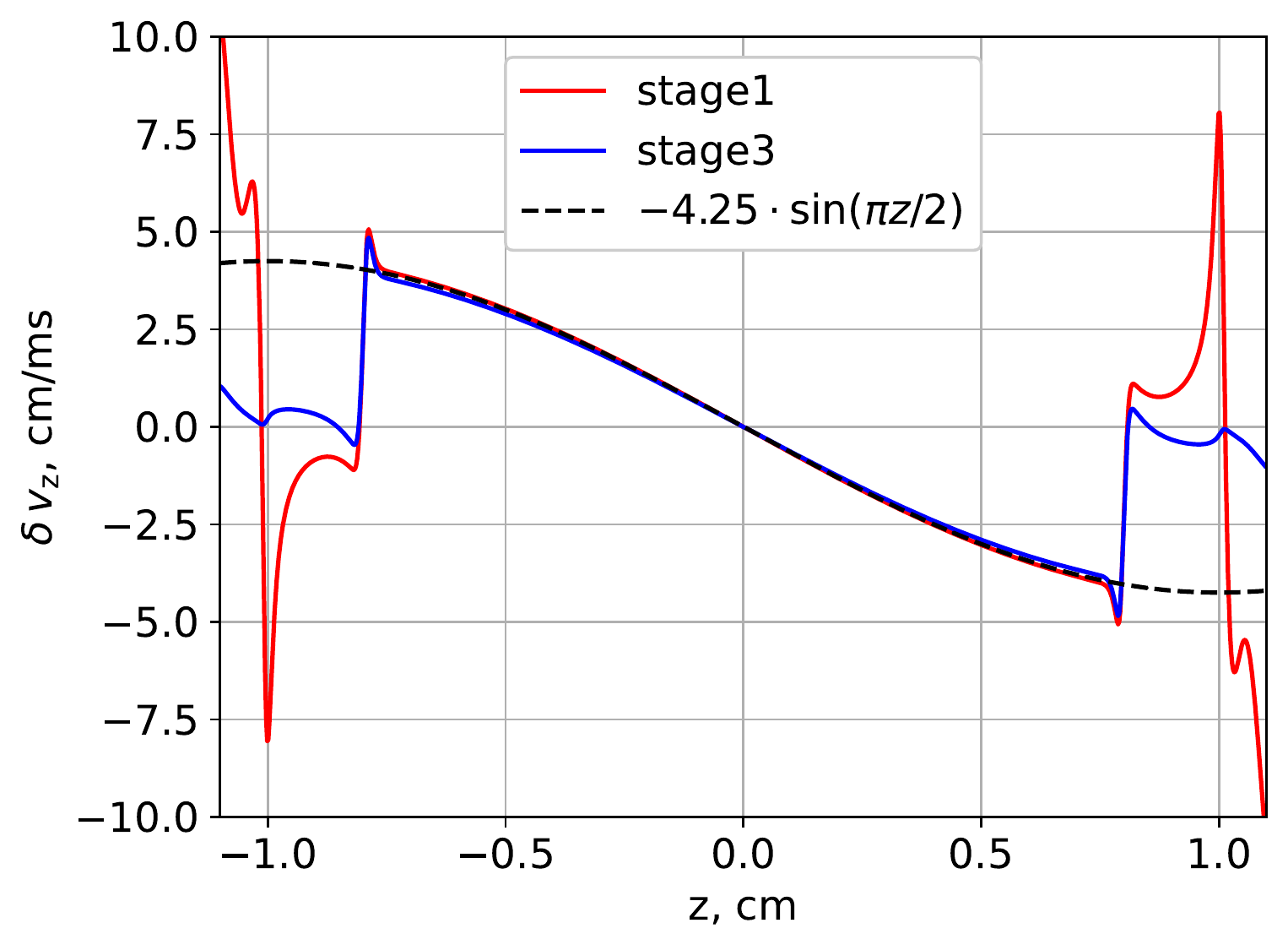}
\caption{ Deviations of parameters of the gas flows from their steady state values for stages 1 and 3 at the moments, when the relative deviations of the gas densities at the capillary center ($r=0$, $z=0$) are equal approximately to 0.05 of their steady state values, $\rho(0,0,t_\infty)$. 
The upper panel shows $\delta(z)=[\rho(0,z,t_{5\%})-\rho(0,z,t=t_\infty)]/\rho(0,0,t_\infty)$ for stages 1 and 3. 
The dashed line shows a fit of the form similar to the theoretical expression~(\ref{R006}).
The lower panel shows simulated axial velocities, $\delta v_{z}(z)=v_z(0,z,t_{5\%})-v_z(0,z,t_\infty)$ for stages 1 and 3. The dashed line shows the fit of the form~\ref{R008}. The theoretical fits in the both panels are relevant only at  $|z| < 0.8$~cm. }
\label{fig10}
\end{figure}

Fig.~\ref{fig8} shows relaxation of the gas density  to steady state at the capillary center($r=0$, $z=0$)  for stages 1 and 3. To compare the plots we shifted  the third stage plot  to the left by 144~$\mu$s. We see that the plots approximately coincide. The difference is greater at the beginning of the stages, when the influence of the initial conditions is more important. 
The plots are approximated by Eqs.~(\ref{Q010}) and ~(\ref{Q020}), when deviation of the central density, $\rho(0,0,t)$, from its steady state value becomes less than 5\%.
It is necessary to note, that simulations of the first and third stages result in almost the same values for $\tau_{
\scriptsize{\mbox{sim}}}$ and $\rho_\infty$.

Thus, the relaxation to the steady state flow develops in the both stages in quite similar ways.  Fig.~\ref{fig10} shows the relative deviations of gas density and axial velocity from their steady-state values on the axis  
for both stages at the moment, when the relative deviations of the gas density and gas velocity along the capillary axis at the capillary center ($r=0$, $z=0$) from their steady state values are 5\%. We see that the distributions of gas density  and velocity along the axis at the axis in  both stages are similar. 

We showed that  the relaxation to the steady state flow does not depend on significant difference in the initial conditions. So we can end the simulation of capillary discharge in repetitive regime at the third stage, because if
we continue  to simulate next  discharge stage and then the recovering of neutral gas distribution  stage we simply repeat the simulation of stages 2 and 3.

\section{The energy deposition by the driver laser pulse}

The femtosecond laser pulse  driving the wake wave deposits energy into the capillary plasma due to  generation of intense plasma wave  and consequent termalization of this energy into internal energy of the electron component. The generation of the wake wave is a key point of the LWFA process, and its parameters are adjusted so that considerable part, $\zeta$, of the laser pulse energy is spent for the plasma wave generation. Typically, this energy deposition is considerably larger than the energy deposited by the electric current pulse, creating the capillary plasma, and much higher than the current capillary plasma energy. As could be assumed that the  dominating energy deposition  leads to considerable change of capillary plasma dynamics. However,   it is not so. Below we explain why we can neglect the effects from the femtosecond laser pulse.

To investigate the influence of  the laser pulse energy deposition on the capillary plasma dynamics we simulate  pulsed energy deposition during the discharge (stage 2). The MHD code cannot simulate  the plasma wave  generation,  damping and energy transformation  into plasma thermal energy. Duration of these processes  ($<0.1$~ns) is much shorter than the duration of the electric current pulse~(\ref{curr}). Then we consider that  on the hydrodynamic time scale duration of these processes is negligible.  We set that an energy deposition takes place at $t=t_\ell$. At this moment of time  electron temperature becomes $T_e=T_\ell$ in the vicinity of the axis, where $r\lesssim r_\ell$. The ion temperature, plasma densities and  velocities are not changed. Then we continue the simulation of stage 2.

 We set for the present simulation: $t_\ell=t_{12}+150$~ns, $T_\ell=500$~eV, and $r_\ell=100~\mu$m. These parameters correspond to $\zeta\simeq 3\times 10^{-4}$ for the 3~J laser pulse.

\begin{figure}[ht] 
\includegraphics[width=0.45\textwidth,clip=
]{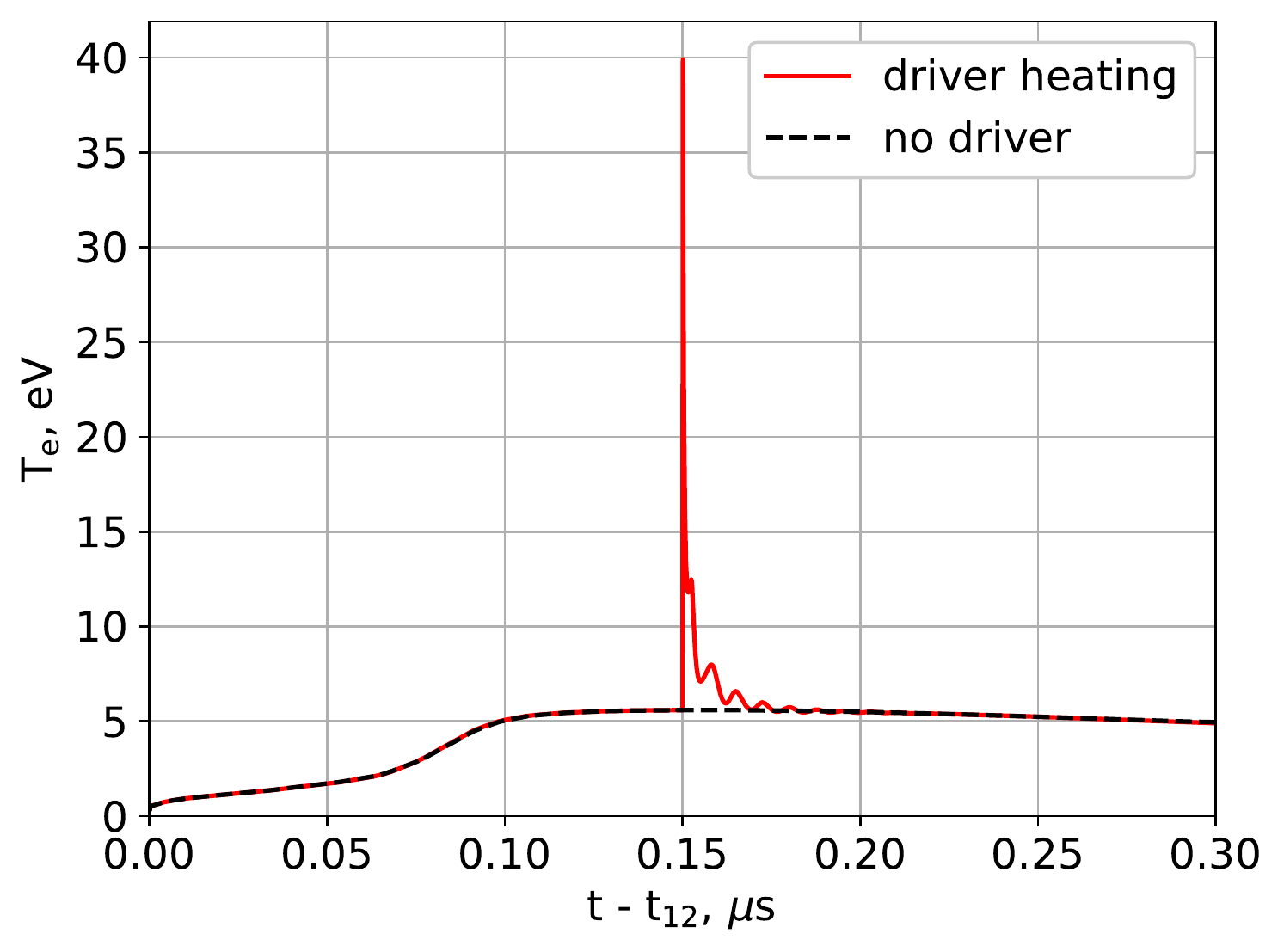} 
\includegraphics[width=0.45\textwidth,clip=
]
{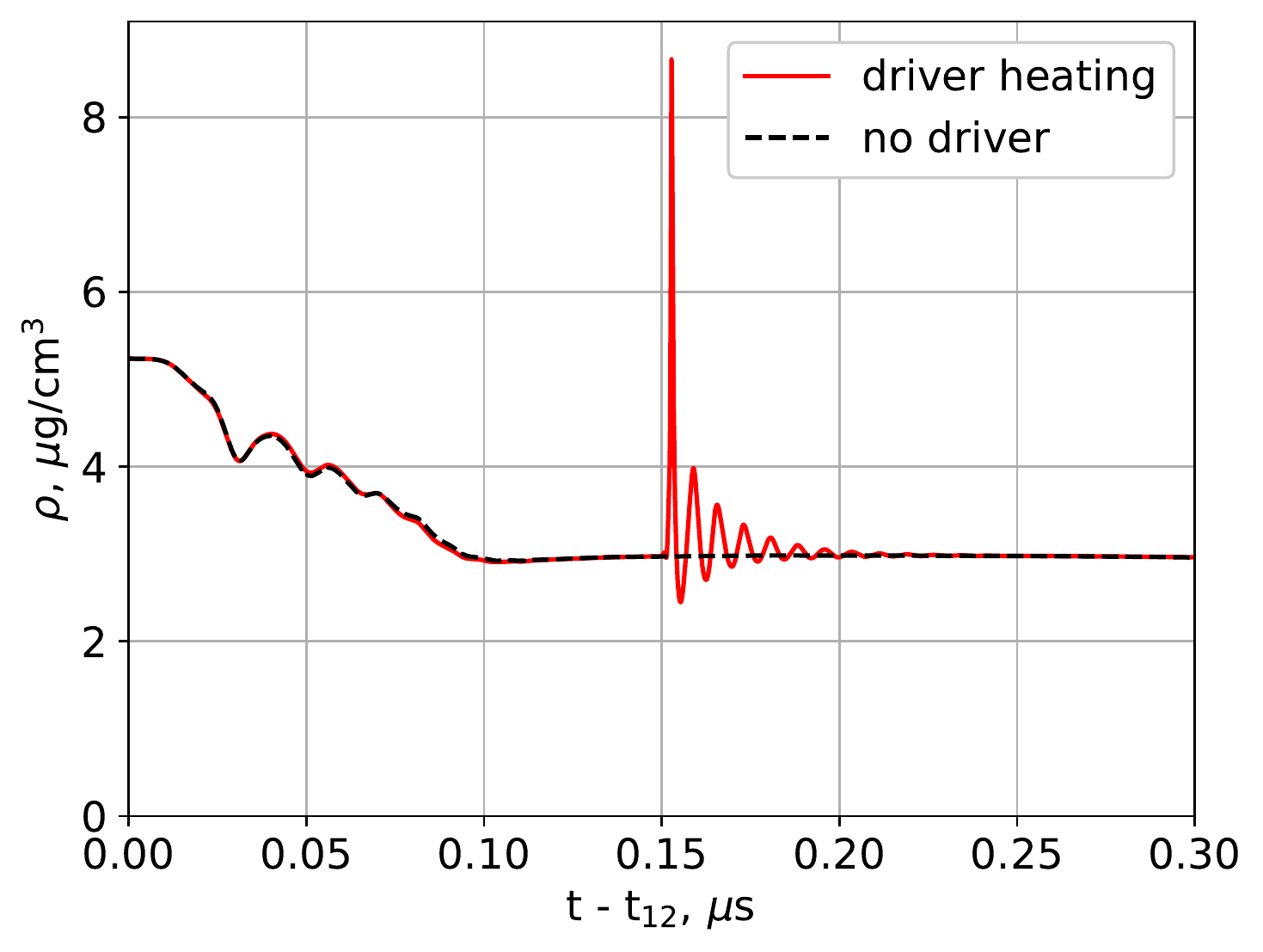}
\caption{Time evolution of the capillary plasma parameters after the driving laser pulse energy deposition. The upper panel shows electron temperature,  the lower panel shows the plasma density at the capillary center ($r=0$, $z=0$). The solid lines  correspond to the simulation with the laser pulse energy deposition, dashed lines to the simulation  without the energy deposition.}
\label{fig11}
\end{figure}

Fig.~\ref{fig11} shows time evolution of the central electron temperature  and plasma density in the cases with and without the femtosecond laser pulse. We see that after approximately  $t\sim15$-20~ns after the laser pulse the state of the capillary plasma returns back to the state without laser energy deposition. Decaying of moderate and then weak oscillations in temperature and plasma density lasts for a longer time, about 30-50~ns after the laser pulse. Such behavior is determined by the plasma dynamics in radial direction.  For the considered capillary discharges the characteristic time of this dynamics  is much shorter  than the electric current pulse characteristic time ~\cite{Bob02}.  A similar problem is discussed in Refs.~\cite{Gon19,Pie20,Bag21}, where additional laser heating of capillary plasma is investigated  to provide more pronounced guiding plasma channel.

\begin{figure*}[ht] 
\includegraphics[width=0.32\textwidth,clip=
]{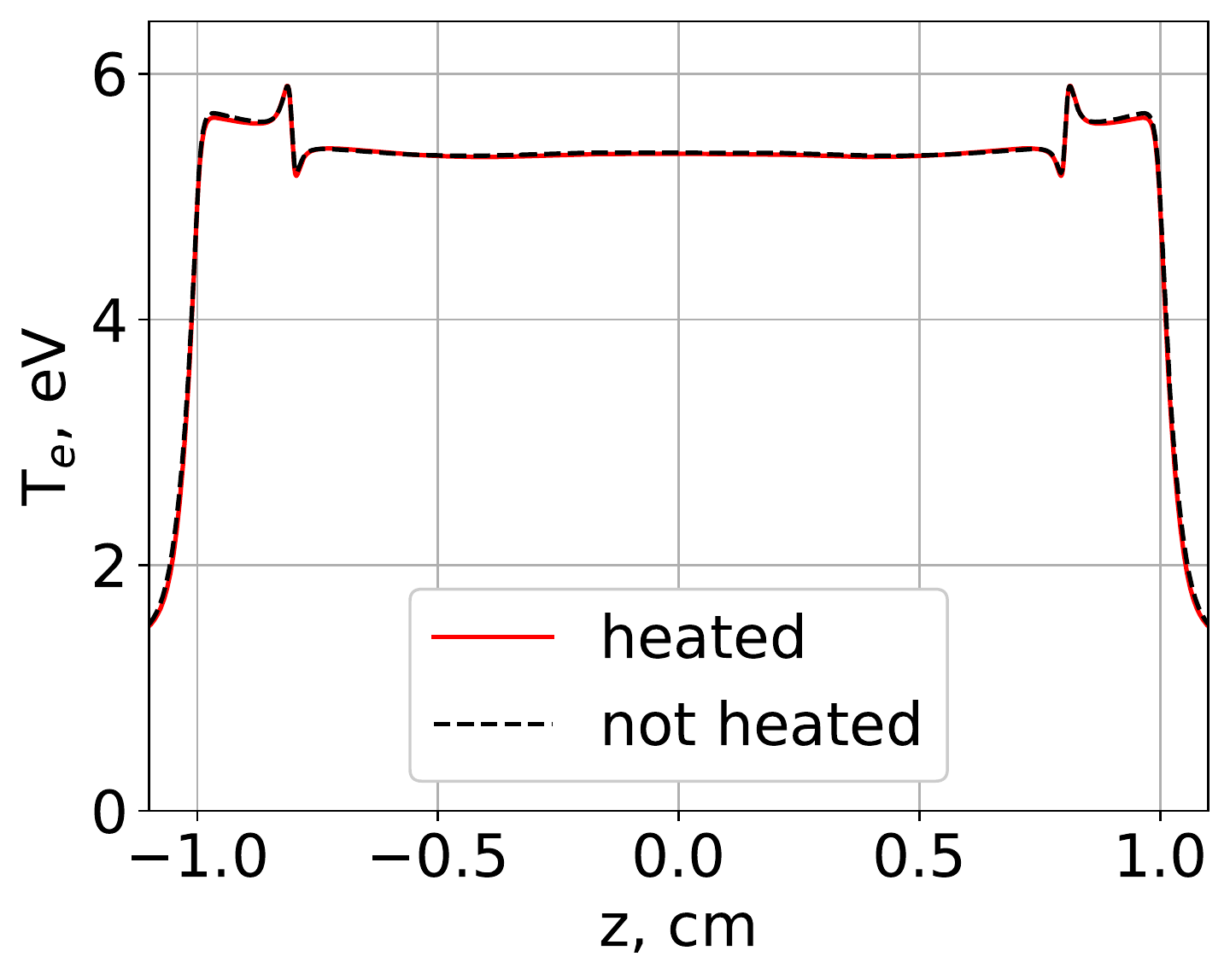} 
\includegraphics[width=0.32\textwidth,clip=
]{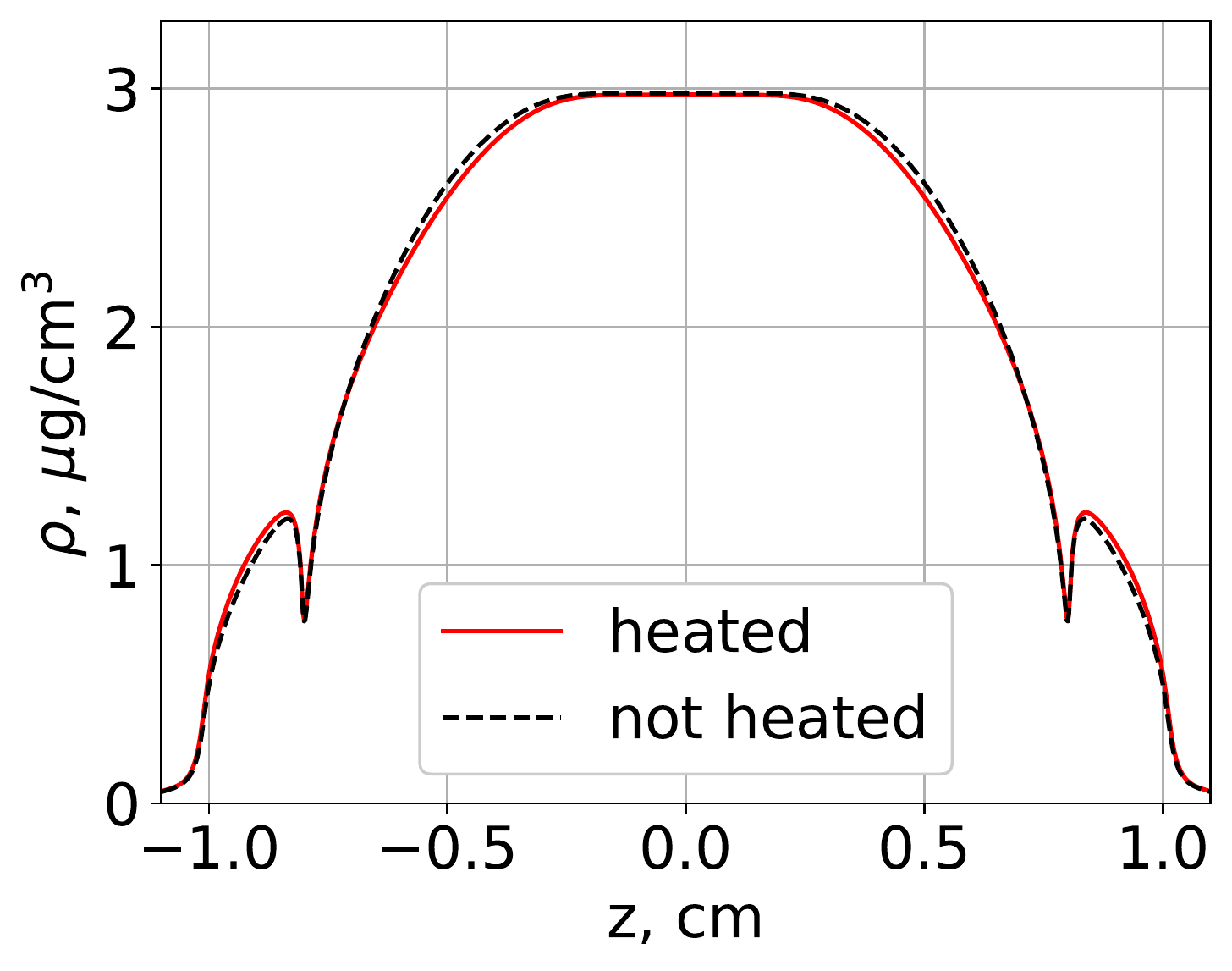}
\includegraphics[width=0.32\textwidth,clip=
]{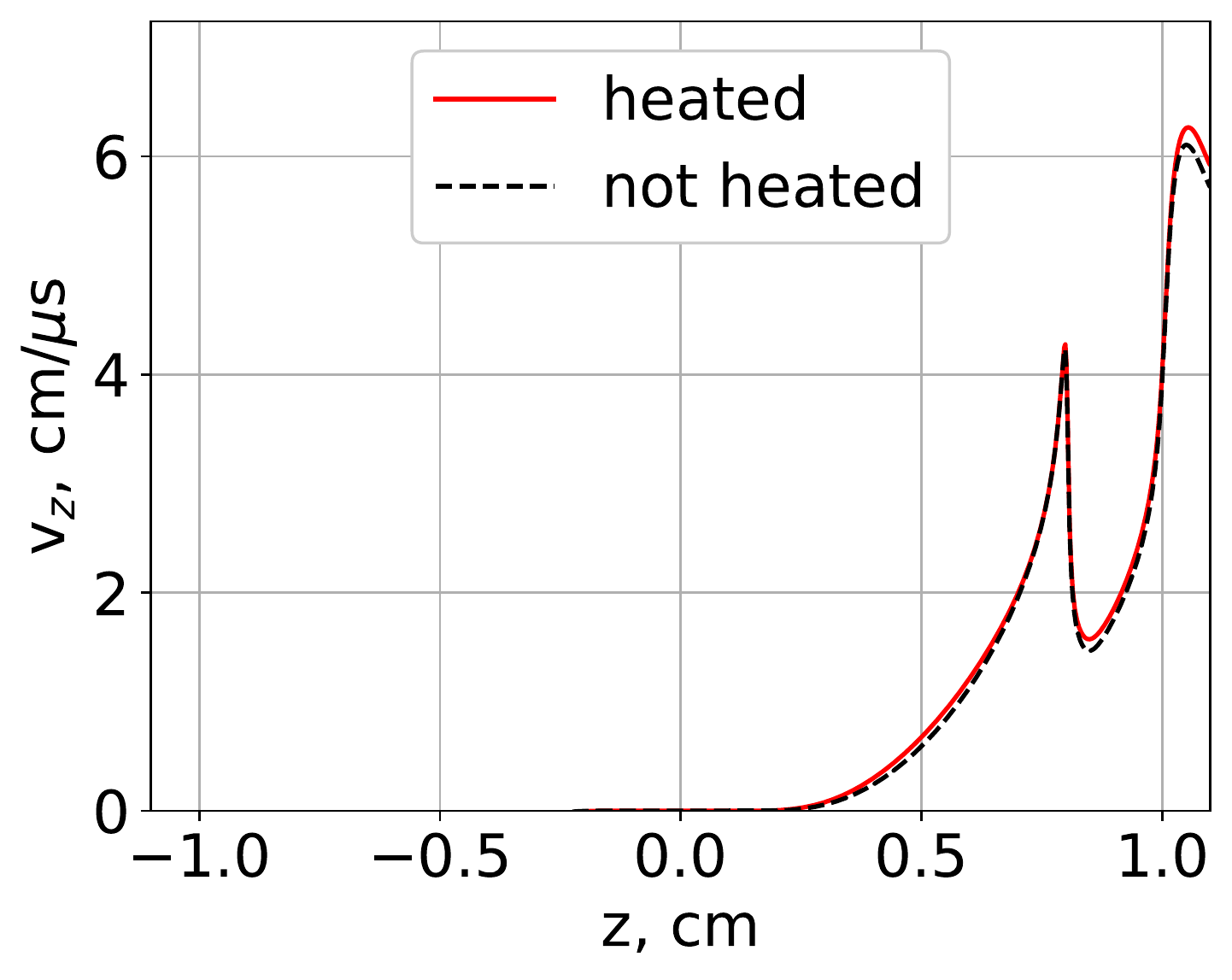}
\caption{Comparison of distributions of the electron temperature, $T_e$, (left panel) plasma density, $\rho$, (central panel) and axial plasma velocity, $v_z$, (right panel) along the capillary  axis at the time $t=t_{12}+230$~ns. The red and black dashed lines correspond to the versions with and without the laser pulse deposition, respectively.
}
\label{fig12}
\end{figure*}

The simulations show (see Fig.~\ref{fig12}), that plasma parameters distributions at $t-t_{12}=230$~ns, remains almost the same as without the laser pulse. The difference between the simulations with and without laser pulse is negligible and we conclude, that  the laser energy deposition have not an effect on the recovery dynamics considered in Secs.~\ref{st3} and~\ref{comp}. The laser energy deposition leads to  minor change of the overall capillary plasma dynamics. Fig.~\ref{fig12} shows that at the time $t=t_{12}+230$~ns  the contribution of the energy deposition to the plasma parameters is less than 2\%. It leads to   0.5\% variation in the recovering time.
 
In  Fig.~\ref{fig11}(upper frame) we do not show the very narrow peak of electron temperature of approximately equal to 500~eV lasting about 0.03~ps. During short time after the energy deposition, when the ions can be considered as immobile, and the electron temperature is governed  by thermal conduction, the electron temperature behaves as $T_e\propto (t-t_\ell)^{-2/5}$, and typical  temperature relaxation time is proportional to $T_e^{-5/2}$. This time becomes too short in the limit of large enough temperature,  to lead to any significant effect on the ion motion. Thus obtained  results remain valid for any very shot time high energy deposition. Hence, when $\zeta\gtrsim 10^{-4}$  considered simulation gives sufficiently accurate description of disturbed plasma dynamics at the time scales $1-20$~ns .

\section{Capillary discharge parameters determining the  repetition rate} 
\label{rep}

In this section we present a theoretical model giving  the neutral hydrogen density distribution recovering time and consider influence of the processes in the capillary plasma on the capillary wall heat balance. In this paper, we do not discuss the process of establishing of a heat balance in the capillary walls~\cite{Gon16}. Nevertheless, we consider here energy deposition from the capillary plasma to the wall.

\subsection{Recovering time of the filling gas steady state distribution  in the capillary}
\label{th}

The process of filling of the capillary with hydrogen, shown in Figs.~\ref{fig3} and~\ref{fig8}, does not posses oscillations. This can be  a consequence of significant role of viscosity in gas dynamics, and  the gas flow can be considered as  Poiseuille flow~\cite{HD}. Then we consider  locally established Poiseuille flow~\cite{HD}  at each cross section of the capillary as a zero-order approximation, neglecting the role of the gas inertia. This yields the following expression for the total mass flow, $q$, through each cross section of the capillary~\cite{HD}: 
 \begin{equation}\label{R002}
 q=-k \frac{S^2\rho}{\eta}\partial_zp=-k \frac{S^2 p}{\eta}\partial_z\rho\, .
 \end{equation}
 Here $S$ is the capillary  cross section area, $\rho$ is the gas density, $\eta$ is the viscosity coefficient, and $p$ is the gas pressure. According to the results, presented in Secs.~\ref{st1}, \ref{st3} and~\ref{comp}, we set  the gas temperature equals to the capillary wall temperature. The gas pressure and density are constant across the cross section. The numerical coefficient, $k$, is of the order of unity. It depends on the form of cross section. For circular form cross section capillary the coefficient is 
 $k=1/8\pi\simeq 0.040$~\cite{HD}.
 We assume that the gas is ideal with constant heat capacity.

Using the mass conservation  ($\partial_t S\rho + \partial_zq=0$), we obtain the  diffusion type equation for the gas density, $\rho$: 
 \begin{equation}\label{R004}
 \partial_t\rho=k\, \partial_z \left(\frac{S p}{\eta}\partial_z\rho\right)\, .
 \end{equation}
 To estimate relaxation time to the steady state flow we consider this equation in a close vicinity of the steady state: $\rho(z,t)=\rho_\infty+\rho_1(z,t)$, ($\rho_1\ll\rho_\infty$). Here $\rho_1$ obeys linear diffusion equation:
 \begin{equation}\label{R0041}
 \partial_t\rho_1=D\partial_z^2\rho_1, 
\end{equation}
where the diffusion coefficient equals $D=kSp/\eta$. The latter linear equation has the lowest mode solution with slowest relaxation rate:  
 \begin{eqnarray}
\rho_1&=&A \rho_\infty e^{-t/\tau} \cos \frac{\pi z}{L}
\, ,\label{R006}\\
 v_z(r=0)&=&
 A\frac{\bar{k}L}{\pi^2\tau}\, e^{-t/\tau}\,\sin \frac{\pi z}{L}
 \, .\label{R008}
 \end{eqnarray}
 Here $\bar{k}=2$  for the circular cross section capillaries,  $L$ is the capillary length, and the relaxation time is
 \begin{equation}\label{R010}
 \tau=\frac{L^2}{\pi^2D}=\frac{1}{k\pi^2}\, \frac{\eta L^2}{p S}\, .
 \end{equation}  
 
 Eqs.~(\ref{R002}) and~(\ref{R010}) are obtained under the conditions: 
\begin{equation}\label{R020} 
\tau\gg \frac{\rho r_0^2}{\eta}\, ,
\quad
\tau\gg\frac{L}{c_s}\, .
\end{equation}
Here, $c_s=\sqrt{p/\rho}$ is isothermal sound velocity of the gas.

For the parameters, presented in Sec.~\ref{setup}, and the hydrogen viscosity coefficient (see ~\cite{Kik}) the right sides of the inequalities ~(\ref{R020}) are equal to 17 and 18~$\mu$s, respectively, whereas Eq.~(\ref{R010}) gives $\tau\simeq 37.4\,\mu$s. When the inequalities~(\ref{R020}) are not strong, the expression~(\ref{R010}) is an estimation by the order of magnitude. The parameter, $\tau$, obtained from simulation (see Secs.~\ref{st1} and~\ref{st3}) is  approximately equal to $\tau_{\scriptsize\mbox{sim}}\simeq 17.4\,\mu$s. The difference between  $\tau$ and $\tau_{\scriptsize\mbox{sim}}$ indicates accuracy of this estimation for the considered parameters~\cite{expl}. For longer capillaries the inequalities are stronger,  then the expression~(\ref{R010})  gives more accurate estimation. Fig.~\ref{fig10} shows simulated profiles  $\rho_1$ and $v_z$ for stages 1 and 3, when the amplitude of $\rho_1$ is about 5\% $\rho_\infty$, and the solutions ~(\ref{R006}) and~(\ref{R008}) for $\rho_1$ and $v_z(r=0)$ at the same moment of time.  $\rho_1\approx A\, \rho_\infty\cos (\pi z/L)$, with $A=-0.05$. For $v_z$ we obtain  $v_z\approx B\sin (\pi z/L)$, with
\begin{equation}\label{R024}
B=\frac{\bar{k} AL}{\pi\tau}\, .
\end{equation}
For $\tau$ equal to the simulated value $\tau_{\scriptsize\mbox{sim}}$, we find that $B\simeq 3.8\times 10^3$~cm/s, whereas the fit in Fig.~\ref{fig10} corresponds to  $B\simeq 4.4\times 10^3$~cm/s. We see that there is reasonable accordance between the simple theory and simulation. It allows us to use the simple theoretical model for the estimations, especially for longer capillaries.
 
In particular if we rewrite the conditions~(\ref{R020}) in terms of the mean free path of hydrogen molecules, $\lambda_{aa}$, both conditions become  equivalent to each other and can be written in the  form:
\begin{equation}\label{R030} 
\frac{L}{r_0}\gg\frac{r_0}{\lambda_{aa}}\, .
\end{equation}
We note that the right hand side of the inequality~(\ref{R030})  very weakly depends on the gas temperature, which is approximately equal to the capillary wall temperature, but significantly depends on the hydrogen density.
 
 After time approximately equal to $\tau_0\sim10\,\mu$s after the discharge, the hydrogen gas density distribution relaxes to a steady state distribution.  Let us compare this statement with Eq.~(\ref{Q020}). The value of $\rho(0,0,\tau_0)$ is about half of its steady state value. At time moment $\Delta t\simeq 4\tau +\tau_0$ after the discharge, the deviation of the density distribution from the steady state distribution is of the order of 1\%.  
It is possible to continue simulation of   the capillary discharge in the repetitive regime with the period $\Delta t$  repeating simulations of stages 2 and 3 in a loop. However it is not necessary, because the deviation of the obtained parameters from the steady state right before the onsets of the electric current pulse is of the order of  1\% and does  not posses any accumulating behavior because the geometry, gas supply pressure and capillary wall temperature remain unchanged in accordance with our assumption.

 We conclude that the repetition rate
\begin{equation}\label{R040} 
f\simeq\frac{1}{C\tau+\tau_0}\, ,
\end{equation}
where $C=4-5\sim\ln 100$, is sufficient for complete recovery of neutral hydrogen distribution before the next current pulse. For the considered parameters the repetition rate of 10~kHz is sufficient for the recovery. If the steady state temperature of the capillary wall is considerably higher than the room temperature, for example, been equal to $1500^\circ$~K, then, according to Eq.~(\ref{R010}), the recovery repetition rate is  two times higher.

\subsection{Energy deposition to the capillary wall}

Here we do not consider  in  details the problems of  the heat balance in the capillary. Here we discuss only how processes in plasma discharge and energy deposition by driver laser pulse  contribute to the heat balance in capillary walls. The repetition rate can be limited by the condition, necessary to maintain the heat balance.

\subsubsection{Energy deposition by the  electric current}
\label{Idep}

Capillary plasma is almost in  thermal and mechanical equilibrium  in the discharges of the type considered in ~\cite{Sp01,Bob02,Sa16}. The details are in Ref.~\cite{Bob02}. We estimate the {\em mean} power, $P$, per unit capillary length deposited to the capillary wall~\cite{Bob02} as:
\begin{equation}\label{E010}
P\left[\mbox{W/cm}\right]=85\left(\frac{I_0[\mbox{kA}]}{r_0[\mbox{mm}]}\right)^{7/5} t_c[\mu\mbox{s}]\, 
f[\mbox{kHz}]\, .
\end{equation}
Here, $t_c$ is the time duration of the electric current pulse (FWHM) with the amplitude $I_0$, and $f$ is the discharge repetition rate. For the considered capillary discharge parameters $P\sim 20$~W/cm for $f=1$~kHz, whereas $P\sim 200$~W/cm for $f=10$~kHz. The total power deposited in the 2~cm length capillary is $W\sim 40$ and 400~W, respectively. To evaluate importance of this effect the latter values are compared with the results of the  heat balance simulations in the capillary walls presented in Fig.~7 of Ref.~\cite{Gon16}.

We considered above the  energy deposition rate averaged over the repetition period. However, the deposited power has narrow peaks during electric current pulses. They cause `{\em pulsed}' heating of a thin layer of the capillary wall in each cycle of the repetitive discharge. Such pulsed heating can damange the capillary after several cycles of the discgarge opreration. This problem was considered in Refs.~\cite{Bob02,Gon16,Vr19}. We conclude that this pulsed effect can be neglected  for the considered parameters of the capillary discharges.

 \subsubsection{Energy deposition by the driver laser pulse}
 
Typical energy of the laser pulse, accelerating electrons to $\sim 0.5$~GeV, is estimated as 3~J~\cite{Mo18}. We assume that a substantial portion, $\zeta$, of this energy is deposited to the capillary plasma is transferred  finally to the plasma internal energy and then  to the capillary wall.   In the experiment~\cite{Gon19}, the coefficient $\zeta$ is approximately equal to 0.5. As a result, total {\em mean} power deposited to the capillary wall owing to this process equals
 $\zeta\times 3$~kW for $f=1$~kHz and $\zeta\times 30$~kW for $f=10$~kHz. We discussed above that 
such energy deposition has a negligible effect on the recovery  time of hydrogen density distribution. Nevertheless, if we compare the latter values with the values presented above in Sec.~\ref{Idep},  we see that the energy deposition by the driver laser pulse is higher than the capillary plasma energy contribution  to the heat balance in the capillary.

If we compare the values presented in the previous paragraph with the results of simulations of heat balance in the capillary walls presented in Fig.~7 of Ref.~\cite{Gon16}, we conclude that laser energy deposited inside the capillary plasma leads to unacceptable heating of the capillary if the repetition rate is of the order of 10~kHz. 
This effect can be mitigated by the use of nitrogen cryogenic cooling~\cite{cooling} and/or capillaries made of diamond~\cite{Gon16}. 

Turning now to the {\em pulsed} energy deposition in the capillary walls, caused by the driver laser pulse, we may say the followings. The high peak temperature of the thin layer of the capillary wall  after the laser pulse during  $\sim100$~ps can lead to melting  or disintegration of the internal capillary wall. This effect is not directly related to the problem of the repetition rate. Nevertheless, it can limit the acceptable range of the driving laser pulse energies, capillary diameters, temperature regimes of the capillary and its materials~\cite{Kam09,Gon19}. Consideration of this important problem is beyond the scope of the present paper and demand a separate work.

\section{Discussion and Conclusions}

 Simulation of one cycle of repetitive operation of  2~cm length and 0.33~mm diameter capillary discharge shows that a plasma channel for laser pulse guiding of the length 0.8~cm and axial electron density about $2\times10^{18}\,\mbox{cn}^{-3}$ exists  for 80~ns, whereas the plasma channel of 1~cm length exists for shorter time, approximately equal to 40~ns. These conditions are favorable for high quality electron beam generation  with the energy of the order of $0.5$~Gev~\cite{Mo18}. For short capillaries the guiding plasma channel lifetime is determined  by the plasma outflow from the open capillary ends and into the supply channels. The electron density in the plasma channel is provided by the stationary hydrogen gas flow $\simeq 0.14\,$mg/s per one supply channel. This requires electric current pulse of duration (FWHM) $\simeq 370\,$ns with the amplitude of 200~A. 
 
The simulation is split into three  stages. The first stage models the process of the capillary filling with a molecular hydrogen gas before the discharge is ignited. The second stage models the capillary discharge till the cooling and recombination of the capillary plasma. The third stage models the recovery of the neutral hydrogen distribution after significant change during the previous stage. We show that a steady state of the gas flow is established at the end of the first and third stages due to permanent supply of hydrogen through the gas inlets. The steady state flow is established with an accuracy better than 1\% after approximately 100~$\mu$s after the electric current peak during the discharge stage. It means that we do not  need to simulate  further cycles of the capillary discharge operation in the repetitive regime  if the  capillary wall temperature, intensity of gas supply and vacuum conditions outside the capillary remain unchanged. If the capillary wall temperature is of the order of the room temperature, the repetition rate can be  $10$~kHz. It provides complete recovering of the hydrogen density distributions between the consequent electric current pulses.

A significant part of driver laser beam energy is deposited into the capillary plasma due to the plasma waves generation. This energy deposition is implemented in the simulation. We show that  the deposited energy is larger than energy deposited into plasma by the electric current, but it causes  negligible change in general gas-plasma dynamics. It happens because a characteristic time of transverse thermal equilibrium establishing is considerably shorter than duration of the electric current pulse for  capillary discharges of this type~\cite{Bob02}.

A simple model of the relaxation time, given by Eqs.~(\ref{R010}), to the steady state flow  based on the use of the relationships of the Poiseuille flow theory is developed in Sec.~\ref{th}. This model is compared with the simulations, what shows  that the simple model gives reasonable  recovery time for the parameters under consideration. 
The model shows that  if the capillary wall temperature is considerably higher than the room temperature, then the recovering time of the neutral hydrogen gas is shorter than the time, obtained in our simulations. For larger capillary lengths the recovering time increases considerably in accordance with Eq.~(\ref{R010}). It can be mitigated by adding extra supply channels between the two existing ones. The hydrogen gas pressure at the inlets of the channels should be less than the pressure for the existing channels to exclude gas flow in the capillary at the steady state.

We estimate  the mean heat flux to the capillary wall from the capillary discharge caused both by  the Joule heating of the capillary plasma and by the energy deposited by the driver laser pulses. 
Although energy deposition by the driver laser pulse leads only to a minor effect on the gas-plasma dynamics in the capillary and, hence, on the recovery time, nevertheless, this repeating energy deposition gives main contribution to the heat flux to the capillary walls. Another important problem related to the energy depositions is damages of the capillary wall after a several cycles of the capillary discharge operation together will driving laser pulses~\cite{Kam09,Gon19}. This important problem is not related  with  the repetition rate and is out of the scope of the present paper.

\begin{acknowledgments}
This research was funded by the project ``Advanced Research using High Intensity Laser produced Photons and Particles'' (ADONIS) (CZ.02.1.01/0.0/0.0/16019/0000789) from European Regional Development Fund (ERDF)  and  by Ministry of Education Youth and Sports of Czech Republic (MEYS CR) grant number CZ.02.1.01/0.0/0.0/16\_019/0000778 .

\end{acknowledgments}


\begin{thebibliography}{99}

\bibitem{Gr07} F.~Grüner, S.~Becker, U.~Schramm, T.~Eichner, M.~Fuchs, R.~Weingartner, D.~Habs, J.~Meyer-ter Vehn, M.~Geissler, M.~Ferrario, L.~Serafini, B.~van der Geer, H.~Backe, W.~Lauth, and S.~Reiche, Design considerations for table-top, laser-based VUV and X-ray free electron lasers, Appl. Phys. B {\bf 86}, 431 (2007).

\bibitem{Fu09} M.~Fuchs, R.~Weingartner, A.~Popp, Z.~Major, S.~Becker, J.~Osterhoff, I.~Cortrie, B.~Zeitler, R.~Hörlein, G.~D.~Tsakiris, U.~Schramm, T.~P.~Rowlands-Rees, S.~M.~Hooker, D.~Habs, F.~Krausz, S.~Karsch, and F.~Grüner, Laser-driven soft-x-ray undulator source, Nat. Phys. {\bf 5}, 826 (2009).

\bibitem{Hu12} Z.~Huang, Y.~Ding, and C.~B.~Schroeder, Compact X-Ray Free-Electron Laser from a Laser-Plasma Accelerator Using a Transverse-Gradient Undulator, Phys. Rev. Lett. {\bf 109}, 204801 (2012).

\bibitem{Co14} M.~E.~Couprie, A.~Loulergue, M.~Labat, R.~Lehe, and V.~Malka, Towards a free electron laser based on laser plasma accelerators, J. Phys. B {\bf 47}, 234001 (2014).

\bibitem{Ma12} A.~R.~Maier, A.~Meseck, S.~Reiche, C.~B.~Schroeder, T.~Seggebrock, and F.~Grüner, Demonstration Scheme for a Laser-Plasma-Driven Free-Electron Laser, Phys. Rev. X {\bf 2}, 031019 (2012).

\bibitem{Mo18} A.~Molodozhentsev, G.~Korn, A.~Maier, and L.~Pribyl, LWFA-driven Free Electron Laser for ELI-Beamlines, in {\em Proceedings of the 60th ICFA Advanced Beam Dynamics Workshop (FLS’18), Shanghai, China, 5-9 March 2018, ICFA Advanced Beam Dynamics Workshop No. 60} (JACoW, Geneva, 2018), pp. 62–67.

\bibitem{As20} R.~W.~Assmann, M.~K.~Weikum, T.~Akhter, D.~Alesini, A.~S.~Alexandrova, M.~P.~Anania, N.~E.~Andreev, I.~Andriyash, M.~Artioli, A.~Aschikhin et al., Eupraxia conceptual design report, Eur. Phys.J.: Spec. Top. {\bf 229}, 3675 (2020).

\bibitem{Ph12} K.~Ta Phuoc, S.~Corde, C.~Thaury, V.~Malka, A.~Tafzi, J.~P.~Goddet, R.~C.~Shah, S.~Sebban, and A.~Rousse, All-optical compton gamma-ray source, Nat. Photonics {\bf 6}, 308 (2012).

\bibitem{Ge15}  C.~G.~R.~Geddes, S.~Rykovanov, N.~H.~Matlis, S.~Steinke, J.-L.~Vay, E.~H.~Esarey, B.~Ludewigt, K.~Nakamura, B.~J.~Quiter, C.~B.~Schroeder, C.~Toth, and W.~P.~Leemans, Compact quasi-monoenergetic photon sources from laser-plasma accelerators for nuclear detection and characterization, Nucl. Instrum. Methods. Phys. Res. B {\bf 350}, 116 (2015).

\bibitem{Kh15} K.~Khrennikov, J.~Wenz, A.~Buck, J.~Xu, M.~Heigoldt, L.~Veisz, and S.~Karsch, Tunable All-Optical Quasimonochromatic Thomson X-Ray Source in the Nonlinear Regime, Phys. Rev. Lett. {\bf 114}, 195003 (2015).

\bibitem{Le09} W.~Leemans and E.~Esarey, Laser-driven plasma-wave electron accelerators, Phys. Today {\bf 62}(3), 44 (2009).

\bibitem{Sch10} C.~B.~Schroeder, E.~Esarey, C.~G.~R.~Geddes, C.~Benedetti, and W.~P.~Leemans, Physics considerations for laser-plasma linear colliders, Phys. Rev. ST: Accel.~Beams {\bf 13}, 101301 (2010).

\bibitem{Fa04} J.~Faure, Y.~Glinec, A.~Pukhov, S.~Kiselev, S.~Gordienko, E.~Lefebvre, J.-P.~Rousseau, F.~Burgy, and V.~Malka, A laser-plasma accelerator producing monoenergetic electron beams, Nature (London) {\bf 431}, 541 (2004).

\bibitem{Ma04} S.~P.~D.~Mangles, C.~D.~Murphy, Z.~Najmudin, A.~G.~R.~Thomas, J.~L.~Collier, A.~E.~Dangor, E.~J.~Divall, P.~S.~Foster, J.~G.~Gallacher, C.~J.~Hooker, D.~A.~Jaroszynski, A.~J.~Langley, W.~B.~Mori, P.~A.~Norreys, F.~S.~Tsung, R.~Viskup, B.~R.~Walton, and K.~Krushelnick, Monoenergetic beams of relativistic electrons from intense laser–plasma interactions, Nature (London) {\bf 431}, 535 (2004).

\bibitem{Os08} J.~Osterhoff, A.~Popp, Z.~Major, B.~Marx, T.~P.~Rowlands-Rees, M.~Fuchs, M.~Geissler, R.~Hörlein, B.~Hidding, S.~Becker, E.~A.~Peralta, U.~Schramm, F.~Grüner, D.~Habs, F.~Krausz, S.~M.~Hooker, and S.~Karsch, Generation of Stable, Low-Divergence Electron Beams by Laser-Wakefield Acceleration in a Steady-State-Flow Gas Cell, Phys. Rev. Lett. {\bf 101}, 085002 (2008).

\bibitem{Li11} J.~S.~Liu, C.~Q.~Xia, W.~T.~Wang, H.~Y.~Lu, C.~Wang, A.~H.~Deng, W.~T.~Li, H.~Zhang, X.~Y.~Liang, Y.~X.~Leng, X.~M.~Lu, C.~Wang, J.~Z.~Wang, K.~Nakajima, R.~X.~Li, and Z.~Z.~Xu, All-Optical Cascaded Laser Wakefield Accelerator Using Ionization-Induced Injection, Phys. Rev. Lett. {\bf 107}, 035001 (2011).

\bibitem{Le14} W.~P.~Leemans, A.~J.~Gonsalves, H.-S.~Mao, K.~Nakamura, C.~Benedetti, C.~B.~Schroeder, C.~Tóth, J.~Daniels, D.~E.~Mittelberger, S.~S.~Bulanov, J.-L.~Vay, C.~G.~R.~Geddes, and E.~Esarey, Multi-Gev Electron Beams from Capillary-Discharge-Guided Subpetawatt Laser Pulses in the Self-Trapping Regime, Phys. Rev. Lett. {\bf 113}, 245002 (2014).

\bibitem{Gon19} A.~J.~Gonsalves, K.~Nakamura, J.~Daniels, C.~Benedetti, C.~Pieronek, T.~C.~H.~de Raadt, S.~Steinke, J.~H.~Bin, S.~S.~Bulanov, J.~van Tilborg, C.~G.~R.~Geddes, C.~B.~Schroeder, Cs.~Tóth, E.~Esarey, K.~Swanson, L.~Fan-Chiang, G.~Bagdasarov, N.~Bobrova, V.~Gasilov, G.~Korn, P.~Sasorov, and W.~P.~Leemans, Petawatt laser guiding and electron beam acceleration to 8 GeV in a laser-heated capillary discharge waveguide,  Phys. Rev. Lett. {\bf 122}, 084801 (2019).

\bibitem{Ta79} Tajima, T. and J. M. Dawson, Laser Electron Accelerator, Phys. Rev. Lett. {\bf 43}, 267 (1979).

\bibitem{Es09} E. Esarey, C. B. Schroeder, and W. P. Leemans, Physics of laser-driven plasma-based electron accelerators, Rev. Mod. Phys. {\bf 81}, 1229 (2009).

\bibitem{Ho00} T.~Hosokai, M.~Kando, H.~Dewa, H.~Kotaki, S.~Kondo, N.~Hasegawa, K.~Nakajima, and K.~Horioka, Optical guidance of terrawatt laser pulses by the implosion phase of a fast z-pinch discharge in a gas-filled capillary, Opt. Lett. {\bf 25}, 10 (2000).

\bibitem{Bu02} A.~Butler, D.~J.~Spence, and S.~M.~Hooker, Guiding of High-Intensity Laser Pulses with a Hydrogen-Filled Capillary Discharge Waveguide, Phys. Rev. Lett. {\bf 89}, 185003 (2002).

\bibitem{Sp03} D.~J.~Spence, A.~Butler, and S.~M.~Hooker, Gas-filled capillary
discharge waveguides, J. Opt. Soc. Am. B {\bf 20}, 138 (2003).


\bibitem{Ka07} S.~Karsch, J.~Osterhoff, A.~Popp, T.~P.~Rowlands-Rees, Z.~Major, M.~Fuchs, B.~Marx, R.~Hörlein, K.~Schmid, L.~Veisz, S.~Becker, U.~Schramm, B.~Hidding, G.~Pretzler, D.~Habs, F.~Grüner, F.~Krausz, and S.~M.~Hooker, GeV-scale electron acceleration in a gas-filled capillary discharge waveguide, New J. Phys. {\bf 9}, 415 (2007).

\bibitem{Kam09} T.~Kameshima, H.~Kotaki, M.~Kando, I.~Daito, K.~Kawase, Y.
Fukuda, L.~M.~Chen, T.~Homma, S.~Kondo, T.~Z.~Esirkepov,
N.~A.~Bobrova, P.~V.~Sasorov, and S.~V.~Bulanov, Laser pulse
guiding and electron acceleration in the ablative capillary discharge
plasma, Phys. Plasmas {\bf 16}, 093101 (2009).

\bibitem{Go11} A.~J.~Gonsalves, K.~Nakamura, C.~Lin, D.~Panasenko, S.
Shiraishi, T.~Sokollik, C.~Benedetti, C.~B.~Schroeder, C.~G.~R.
Geddes, J.~van Tilborg, J.~Osterhoff, E.~Esarey, C.~Toth, and
W.~P.~Leemans, Tunable laser plasma accelerator based on longitudinal
density tailoring, Nat. Phys. {\bf 7}, 862 (2011).


\bibitem{Pa50} W.~K.~H.~Panofsky and W.~R.~Baker, A focusing device for the external 350-MeV proton beam of the 184-inch cyclotron at Berkeley, Rev. Sci. Instrum. {\bf 21}, 445 (1950).

\bibitem{Ti15} J.~van Tilborg, S.~Steinke, C.~G.~R.~Geddes, N.~H.~Matlis, B.~H.~Shaw, A.~J.~Gonsalves, J.~V.~Huijts, K.~Nakamura, J.~Daniels, C.~B.~Schroeder, C.~Benedetti, E.~Esarey, S.~S.~Bulanov, N.~A.~Bobrova, P.~V.~Sasorov, and W.~P.~Leemans, Active Plasma Lensing for Relativistic Laser-Plasma-Accelerated Electron Beams, Phys. Rev. Lett. {\bf 115}, 184802 (2015).

\bibitem{Po19} A.~Ferran Pousa, A.~Martinez de la Ossa, R.~Brinkmann, and R.~W.~Assmann, Compact Multistage Plasma-Based Accelerator Design for Correlated Energy Spread Compensation, Phys. Rev. Lett. {\bf 123}, 054801 (2019).

\bibitem{Eh96} Y.~Ehrlich, C.~Cohen, A.~Zigler, J.~Krall, P.~Sprangle, and E.~Esarey, Guiding of High Intensity Laser Pulses in Straight and Curved Plasma Channel Experiments, Phys. Rev. Lett. {\bf 77}, 4186 (1996).

\bibitem{Sp01} D.~J.~Spence and S.~M.~Hooker, Investigation of a hydrogen
plasma waveguide, Phys. Rev. E 63, 015401(R) (2001).

\bibitem{Bob02}N.~A.~Bobrova, A.~A.~Esaulov, J.-I.~Sakai, P.~V.~Sasorov, D.~J.~Spence, A.~Butler, S.~M.~Hooker, S.~V.~Bulanov, Simulations of a hydrogen-filled capillary discharge waveguide, Phys. Rev. E {\bf 65}, 016407 (2002).
\bibitem{Pie20} C.~Pieronek, A.~Gonsalves, C.~Benedetti, S.~Bulanov, J.~van Tilborg, J.~Bin, K.~Swanson, J.~Daniels, G.~Bagdasarov, N.~Bobrova, V.~Gasilov, G.~Korn, P.~Sasorov, C.~Geddes, C.~Schroeder, W.~Leemans, and E.~Esarey, Laser-heated capillary discharge waveguides as tunable structures for laser-plasma acceleration, Phys. Plasmas {\bf 27}, 093101 (2020).
\bibitem{Bob13} N.~A.~Bobrova, P.~V.~Sasorov, C.~Benedetti, S.~S.~Bulanov, C.~G.~R.~Geddes, C.~B.~Schroeder, E.~Esarey, and W.~P.~Leemans, Laser-heater assisted plasma channel formation in capillary discharge waveguides, Phys. Plasmas {\bf 20}, 020703 (2013).
\bibitem{Bag21} G.~A.~Bagdasarov, N.~A.~Bobrova, O.~G.~Olkhovskaya, V.~A.~Gasilov, C.~Benedetti, S.~S.~Bulanov, A.~J.~Gonsalves, C.~V.~Pieronek, J.~van Tilborg, C.~G.~R.~Geddes, C.~B.~Schroeder, P.~V.~Sasorov, S.~V.~Bulanov, G.~Korn, and E.~Esarey, Creation of axially uniform plasma channel in laser-assisted capillary discharge, Phys. Plasmas {\bf 28}, 053104 (2021).

\bibitem{Gon16} A.~J.~Gonsalves, F.~Liu, N.~A.~Bobrova, P.~V.~Sasorov, C.~Pieronek, J.~Daniels, S.~Antipov, J.~E.~Butler, S.~S.~Bulanov, W.~L.~Waldron, D.~E.~Mittelberger, and W.~P.~Leemans, Demonstration of a High Repetition Rate Capillary Discharge Waveguide, J.~Appl.~Phys.~{\bf 119}, 033302 (2016).
\bibitem{Al22} A.~Alejo, J.~Cowley, A.~Picksley, R.~Walczak, and S.~M.~Hooker, Demonstration of kilohertz operation of hydrodynamic optical-field-ionized plasma channels, Phys.~Rev.~Accel.~Beams {\bf 25}, 011301 (2022).
\bibitem{Arc22} R.~D’Arcy, J.~Chappell, J.~Beinortaite, S.~Diederichs, G.~Boyle, B.~Foster, M.~J.~Garland,
P.~Gonzalez Caminal, C.~A.~Lindstrøm, G.~Loisch, S.~Schreiber, S.~Schröder, R.~J.~Shalloo, M.~Thévenet, S.~Wesch, M.~Wing, and J.~Osterhoff, Recovery time of a plasma-wakefield accelerator, Nature {\bf 603}, 58 (2022).

\bibitem{MS22} B.~Miao, J.~E.~Shrock,  L.~Feder, R.C.~Hollinger, J.~Morrison, R.~Nedbailo, A.~Picksley, H.~Song, S.~Wang, J.~J.~Rocca, and H .~M.~Milchberg, Phys. Rev. X {\bf 12}, 031038 (2022).

\bibitem{Ga12} V.~Gasilov, A.~Boldarev, S.~Dyachenko, O.~Olkhovskaya, E.~Kartasheva, G.~Bagdasarov, S.~Boldyrev, I.~Gasilova, V.~Shmyrov, S.~Tkachenko, J.~Grunenwald, and T.~Maillard, Towards an application of high-performance computer systems to 3D simulations of high energy density plasmas in Z-Pinches, in Applications, Tools and Techniques on the Road to Exascale Computing, Advances in Parallel Computing Vol.~22 (IOS Press BV, Amsterdam, 2012), pp.~235–242.

\bibitem{Ol20} O.~G.~Olkhovskaya, G.~A.~Bagdasarov, N.~A.~Bobrova,,  V.~A.~Gasilov, L.~V.~N.~Goncalves, 
C.~Lazzarini, M.~Nevrkla,  G.~Grittani, S.~S.~Bulanov, A.~G.~Gonsalves, C.~B.~Schroeder, E.~Esarey, P.~V.~Sasorov,  S.~V.~Bulanov,  and G.~Korn, Plasma channel formation in the knife-like focus of laser beam, J. Plasma Phys. {\bf 86}, 905860307 (2020).

\bibitem{Bag22} G.~A.~Bagdasarov, K.~O.~Kruchinin, A.~Yu.~Molodozhentsev, P.~V.~Sasorov, S.~V.~Bulanov, V.~A.~Gasilov, Discharge Plasma Formation in Square Capillary with Gas Supply Channels, Phys. Rev. Res. {\bf 4}, 013063 (2022).

\bibitem{Bag17} G.~Bagdasarov, P.~Sasorov, V.~Gasilov, A.~Boldarev, O.~Olkhovskaya, C.~Benedetti, S.~Bulanov, A.~Gonsalves, H.~S.~Mao, C.~B.~Schroeder, J.~van Tilborg, E.~Esarey, W.~Leemans, T.~Levato, D.~Margarone, and G.~Korn, Laser beam coupling with capillary discharge plasma for laser wakefield acceleration applications, Phys. Plasmas {\bf 24}, 083109 (2017).

\bibitem{Gr08} E.~V.~Grabovski, V.~V.~Aleksandrov, G.~S.~Volkov, V.~A.~Gasilov, A.~N.~Gribov, A.~N.~Gritsuk, S.~V.~Dyachenko, V.~I.~Zaitsev, S.~F.~Medovshchikov, K.~N.~Mitrofanov, Ya.~N.~Laukhin, G.~M.~Oleinik, O.~G.~Olkhovskaja, A.~A.~Samokhin, P.~V.~Sasorov, and I.~N.~Frolov, Use of Conical Wire Arrays for Modeling Three-Dimensional MHD Implosion Effects, Plasma Phys. Reps. {\bf 34}, 815 (2008).

\bibitem{Ale19} V.~Aleksandrov, A.~Branitski, V.~Gasilov, E.~Grabovskiy, A.~Gritsuk, K.~Mitrofanov, O.~Olkhovskaya P.~Sasorov, I.~Frolov, Study of interaction between plasma flows and the magnetic field at the implosion of nested wire arrays, Pl. Phys. Contr. Fusion {\bf 61}, 035009 (2019).

\bibitem{HD} L.~D.~Landau and E.~M.~Lifshitz,  {\em Fluid Mechanics} (Oxford: Reed) 2000.

\bibitem{Kik} I.~S.~Grigoriev, and E.~Z.~Meilikhov, {\em Handbook of Physical Quantities} (CRC Press, Boca Raton, FL, 1997).

\bibitem{expl} Considering a damped oscillator ($\ddot{x}+2\nu\dot{x}+\omega_0^2x=0$), we have for the over-damped case: $\tau\approx 2\nu\omega_0^{-2}$, whereas for critical damping ($\nu=\omega_0$), we have $\tau=\omega_0$ that is two times longer than the over-damped limit predicts. This argument helps to understand that when the both sides of the 2nd strong inequality of Eq.~(\ref{R020}) become of the same order, then we may be close to the critically damped regime and the parameter $\tau$ can be in exreme case about two times less than the estimation~(\ref{R010}), coming from the Poiseuille approximation, based on  neglecting of gas inertia, gives.

\bibitem{Sa16} P. Sasorov, Plasma Dynamics in Capillary Discharges, in ``X-ray lasers 2016'', Proceedings of the 15th International Conference on X-Ray Lasers, Springer Proceedings in Physics Vol. 202, edited by T. Kawachi, S. V. Bulanov, H. Daido, and Y. Kato, 2018, p. 45.

\bibitem{Vr19} M.~Vrbova, P.~Vrba, A.~Jancarek, M.~Nevrkla, N.~A.~Bobrova, and P.~V.~Sasorov, Wall ablation effect on the recombination pumping of EUV laser in  pinching capillary discharge, Phys. Plasmas {\bf 26}, 083108 (2019).



\bibitem{Kin} E.~M.~Lifshitz and L.~P.~Pitaevskii, {\em Physical Kinetics} (Pergamon, New York, 1981).

\bibitem{cooling} Cooling of sufficiently pure and perfect crystal dielectrics leads to increasing of thermal conductivity coefficient~\cite{Kin} in the capillary wall and, hence, to decreasing the capillary internal wall temperature.


\end{thebibliography}
\end{document}